\input{aipcheck}

\documentclass[
  ,draft            % use draft while you are working on the paper
  ]
  {aipproc}

\layoutstyle{6x9}

\newcommand{\rd}{{\rm{d}}}
\newcommand{\vphi}{\varphi}
\newcommand{\vepsilon}{\varepsilon}

\newcommand{\sqdetg}{\sqrt{-g}}

\begin{document}

\title{Rotating Black Holes in Higher Dimensions}


\author{Burkhard Kleihaus}{
  address={ZARM, Universit\"at Bremen, Am Fallturm,
D--28359 Bremen, Germany}
%Institut f\"ur Physik, Universit\"at Oldenburg, Postfach 2503\\
%D-26111 Oldenburg, Germany}
}

\author{Jutta Kunz}{
  address={Institut f\"ur Physik, Universit\"at Oldenburg, Postfach 2503,
D-26111 Oldenburg, Germany}
}

\author{Francisco Navarro-L\'erida}{
  address={Dept.~de F\'{\i}sica At\'omica, Molecular y Nuclear, Ciencias F\'{\i}sicas\\
Universidad Complutense de Madrid, E-28040 Madrid, Spain}
  ,altaddress={Institut f\"ur Physik, Universit\"at Oldenburg, Postfach 2503,
D-26111 Oldenburg, Germany} % additional visiting address
}

\begin{abstract}
The properties of higher-dimensional black holes can differ
significantly from those of black holes in four dimensions,
since neither the uniqueness theorem, nor the staticity theorem
or the topological censorship theorem generalize to higher
dimensions.
We first discuss black holes of Einstein-Maxwell theory
and Einstein-Maxwell-Chern-Simons theory
with spherical horizon topology.
Here new types of stationary black holes are encountered.
We then discuss nonuniform black strings and present evidence for
a horizon topology changing transition.
\end{abstract}

\maketitle

%%%%%%%%%%%%%%%%%%%%%%%%%%%%%%%%%%%%%%%%%%%%
%% MAINMATTER
%%%%%%%%%%%%%%%%%%%%%%%%%%%%%%%%%%%%%%%%%%%%

\section{Introduction}

Black holes are a major prediction of Einstein's general relativity.
Today there is strong observational
evidence for the existence of astrophysical black holes.
On the other hand string theory,
a major candidate for the quantum theory of gravity
and the unification of all interactions, predicts in its low energy
limit additional fields and also requires higher dimensions for mathematical
consistency. As a result, essential properties of black holes can change
dramatically.
Here we address the questions
as to what the consequences for the properties of black holes are
and how they are affected by the presence of extra dimensions.

Black holes in 4-dimensional Einstein-Maxwell (EM) theory 
have a number of important special properties.
First of all, they are subject to the
topological censorship theorem, stating that their horizons
have the topology of a sphere, $S^2$ \cite{Hawking:1973uf,Friedman:1993ty}.
Then they satisfy a uniqueness theorem,
stating they are uniquely characterized
by their global charges:
their mass $M$, their angular momentum $J$,
their electric charge $Q$, and their magnetic charge $P$
\cite{Israel:1967za,Robinson:1975bv,Mazur:1982db,Heusler:1996}.
They also obey the staticity theorem,
stating that stationary black holes with static (non-rotating) horizons
must be static, i.e., they carry no angular momentum $J$ \cite{Wald:1993ki}.

\begin{center}
\begin{tabular}{|c|c|c|} \hline
 & static & rotating \\\hline
uncharged & Schwarzschild & Kerr \\
 &  ($M$) &   ($M,J$) \\ \hline
charged & Reissner-Nordstr\"om & Kerr-Newman \\
 & ($M,Q,P$) & ($M,Q,P,J$) \\ \hline
\end{tabular}
\end{center}

EM black holes further satisfy the laws of black hole mechanics.
According to the zeroth law,
their temperature $T$ is constant on their horizon,
where their temperature is proportional to their surface gravity $\kappa$,
$T = {\kappa}/{(2 \pi)}$.
The first law reads
\begin{equation}
\rd M= \frac{\kappa}{8\pi G} \rd {A_{\rm H}} + {\Omega}  \rd {J}
+ \Phi_{\rm H} \rd Q \, ,
\label{first4}
\end{equation}
where the black hole horizon area ${A_{\rm H}}$ is proportional to 
the entropy $S$, $S= {A_{\rm H}}/(4 G)$,
$\Omega$ denotes the horizon angular velocity,
and $\Phi_{\rm H}$ 
represents the horizon electrostatic potential 
(and in the presence of magnetic charge a further term enters).
The integrated first law then yields the Smarr formula \cite{Smarr:1972kt},
relating horizon properties and global charges,
\begin{equation}
M = \frac{2}{8 \pi G} \kappa {A_{\rm H}} + 2 \Omega J + \Phi_{\rm H} Q
\ . \label{smarr4} \end{equation}

In $D=4$ dimensions the Kerr black holes satisfy the relation
\hbox{$M^2\ge \mid 16\pi J\mid$}, while for the Kerr-Newman black holes
of EM theory $M^2\ge 4 Q^2+(16 \pi J)^2/M^2$ holds. 
Thus the angular momentum of $D=4$ EM black holes is always
bounded from above.
For extremal solutions the bounds are precisely saturated.
Therefore extremal solutions
enclose the domain of existence of EM black holes.
Solutions with greater angular momenta do not possess a horizon.
Instead they exhibit a naked singularity, violating cosmic censorship.

The generalization of these asymptotically flat 
black hole solutions with spherical horizon topology
to $D>4$ dimensions was pioneered by Tangherlini
\cite{Tangherlini:1963bw} for static EM black holes,
and by Myers and Perry (MP) \cite{Myers:1986un} 
for rotating vacuum black holes.
Stationary black holes in $D$ dimensions
possess $N=[(D-1)/2]$ independent angular momenta $J_i$
associated with $N$ orthogonal planes of rotation \cite{Myers:1986un}.
($[(D-1)/2]$ denotes the integer part of $(D-1)/2$, corresponding to the
rank of the rotation group $SO(D-1)$.)
The general black hole solutions then fall into two classes,
in even-$D$ and odd-$D$ solutions \cite{Myers:1986un}.

\begin{figure}[h!]
\parbox{\textwidth}
{\centerline{
\mbox{
\includegraphics[width=60mm,angle=0,keepaspectratio]{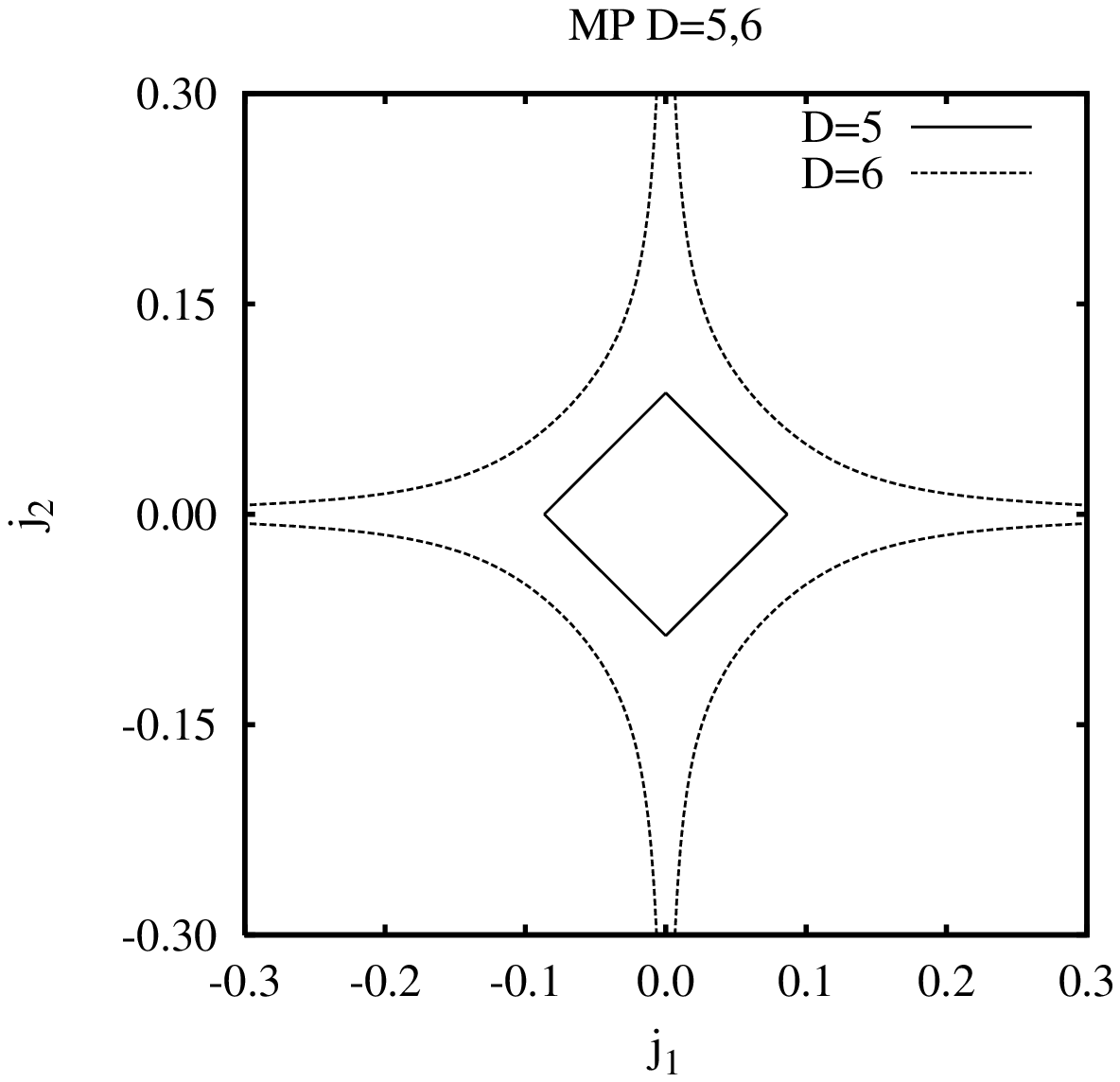}
\includegraphics[width=90mm,angle=0,keepaspectratio]{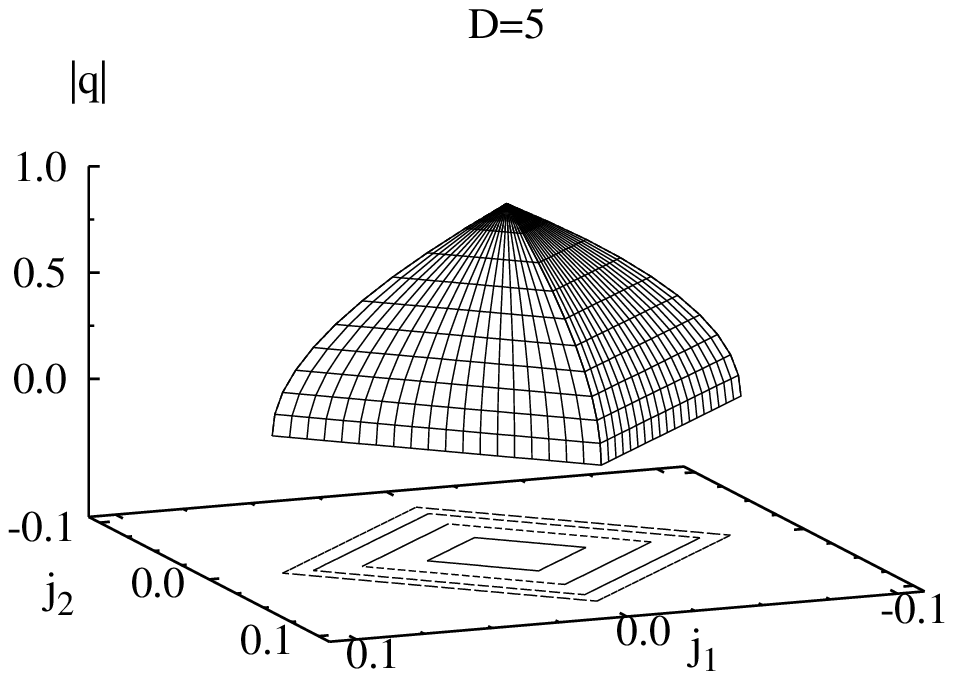}
}}}
\caption{
Left:
Domain of existence of MP black holes in $D=5$ and $D=6$:
scaled angular momentum $j_1=J_1/M^{(D-2)/(D-3)}$ versus
$j_2=J_2/M^{(D-2)/(D-3)}$
for extremal solutions.
Right:
Domain of existence of EMD black holes in $D=5$:
scaled charge $|q|=|Q|/M$
versus scaled angular momenta $j_i=J_i/M^{(D-2)/(D-3)}$, $i=1,2$
for extremal EMD solutions.
}
\label{f-1}
\end{figure}

In $D=5$ and $D=6$ dimensions the MP black holes 
then possess two independent angular momenta $J_i$, $i=1,\, 2$.
Introducing the scaled angular momenta $j_1=J_i/M^{(D-2)/(D-3)}$,
we exhibit the domain of existence for these MP black holes
in Fig.~\ref{f-1}.
The extremal $D=5$ MP solutions then
form a square with respect to the scaled angular momenta,
thus in 5 dimensions the angular momenta of these MP black holes
are always bounded from above.
(At the vertices of the square
one of the two angular momenta vanishes,
and the associated extremal single angular momentum solutions
possess vanishing horizon area.)

When moving to $D=6$ dimensions,
the domain of existence of MP solutions changes distinctly.
The vertices present in the $D=5$ domain then move to infinity.
Thus there are no extremal solutions
for single angular momentum black holes.
This holds also for black holes in more than $6$ dimensions.
As a consequence of this unlimited growth of the angular momentum
instabilities develop, where new branches of black holes
are expected to arise (as alluded to in the conclusions) 
\cite{Emparan:2003sy,Emparan:2007wm}.

The corresponding $D>4$ charged rotating black holes of pure EM theory
could not yet be obtained in closed form.
In contrast to pure EM theory, exact solutions of higher dimensional
charged rotating black holes are known in theories with more symmetries.
The inclusion of additional fields, as required by supersymmetry
or string theory, leads to further exact solutions of higher
dimensional black holes \cite{Horowitz:1995tm,Youm:1997hw},
since then certain constructive methods are available, such as
Hassan-Sen transformations \cite{Hassan:1991mq,Sen:1994eb}

Einstein-Maxwell-dilaton (EMD) black holes, for instance, are obtained
by embedding the $D$-dimensional Myers-Perry solutions
in $D+1$ dimensions, and performing a boost with respect to
the time and the additional coordinate, followed by a
Kaluza-Klein reduction to $D$ dimensions 
\cite{Llatas:1996gh,Kunz:2006jd}.
The domain of existence of such $D=5$ EMD black holes is also shown
in Fig.~\ref{f-1}.

Here we first discuss charged rotating black holes in pure
EM theory, obtained by numerical as well as perturbative methods.
We focus on rotating black holes in odd dimensions,
whose $N$ angular momenta have all equal-magnitude.
The reason is, that because of symmetry
the solutions then greatly simplify:
The general MP solutions with $N$ independent angular momenta $J_i$
possess $U(1)^N$ symmetry.
For odd-$D$ black holes the symmetry is then strongly enhanced
(to a $U(N)$ symmetry),
when all $N$ angular momenta have equal-magnitude,
and consequently the angular dependence factorizes.
This factorization of the angular dependence
also occurs in the presence of a gauge field,
and thus in EM theory.

In odd dimensions $D=2N+1$ the Einstein-Maxwell action
may be supplemented by a `$A\,F^N$' Chern-Simons (CS) term.
The bosonic sector of minimal $D=5$ supergravity
may be viewed as the special $\lambda=1$ case
of the general Einstein-Maxwell-Chern-Simons (EMCS)
theory with Lagrangian
\begin{equation}
{\cal L}= \frac{1}{16\pi G_5} \left[\sqrt{-g}(R -F^2) -
\frac{2\lambda}{3\sqrt{3}}\varepsilon^{mnpqr}A_mF_{np}F_{qr}\right] \ ,
\label{Lag}
\end{equation}
and CS coefficient $\lambda$.
While not affecting the static black hole solutions,
the CS term does affect the stationary black hole solutions.
The addition of the CS term
makes it easier to solve the field equations
in the special case of the supergravity coefficient $\lambda=1$,
and analytic solutions describing charged, rotating black holes
are known 
\cite{Breckenridge:1996sn,Breckenridge:1996is,Cvetic:2004hs,Chong:2005hr}.

We here discuss the properties
of $D=5$ black holes in general EMCS theories,
treating the CS coefficient as a parameter.
Beyond the supergravity value, i.e., beyond $\lambda = 1$,
we find new types of stationary black hole solutions,
such as counterrotating black holes, where the
horizon angular velocity and the angular momentum have opposite
sense, or regular (non extremal)
black holes which possess a static horizon but finite
angular momenta.
Beyond $\lambda = 2$ we then observe further new phenomena,
such as nonuniqueness of black holes with spherical horizon topology
and black holes with negative horizon mass.
Thus neither the uniqueness theorem nor the staticity theorem
generalize to higher dimensions.

Indeed, in recent years it has been realized that
black holes exhibit a much richer structure in higher dimensions
than in four dimensions.
In particular, black objects with different types of horizon
topologies are present in higher dimensions,
since extensions of the topological censorship theorems
put very little constraints on the horizon topology in higher
dimensions.
Black rings \cite{Emparan:2001wn,Emparan:2006mm}
in $D=5$ dimensions, for instance, possess the horizon
topology of a torus, $S^1 \times S^2$, and there are also
concentric black rings or black saturns 
\cite{Gauntlett:2004wh,Elvang:2007rd}.
All these black objects are asymptotically flat.

But the higher dimensions may also be compact, and the 
corresponding black objects will then not be asymptotically flat.
Assuming one dimension to be compact, caged black holes appear,
as found in five and six dimensions
\cite{Kol:2003if,Sorkin:2003ka,Kudoh:2003ki,Harmark:2003yz}.
For small $S^{D-2}$ horizons these caged black holes differ only little
from asymptotically flat black holes. For larger black holes, however,
the compact dimension becomes essential, since at a critical size,
the horizon will cover the compact dimension completely.
The horizon topology must then change from $S^{D-2}$
to $S^{D-3}\times S^1$, i.e., a horizon topology changing transition is
expected \cite{Kol:2002xz,Kol:2004ww,Harmark:2005pp,Harmark:2007md}.

%\vspace{-1cm}
\begin{figure}[h!]
\setlength{\unitlength}{1cm}
\begin{picture}(19,3)
%\hspace{-0.5cm}
\put(0,-1){
\includegraphics[width=75mm,angle=0,keepaspectratio]{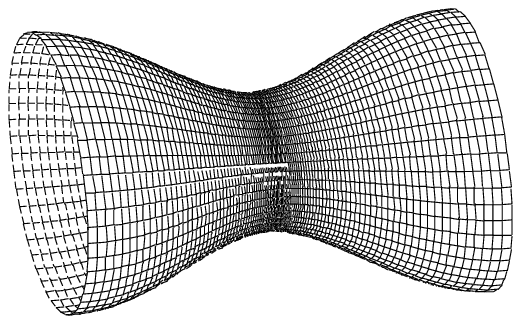}
}
%\hspace{-2.5cm}
\put(5.0,-1){
\includegraphics[width=75mm,angle=0,keepaspectratio]{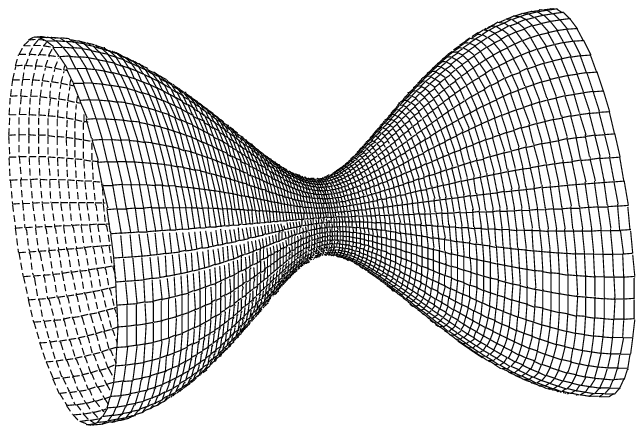}
}
%\hspace{-2cm}
\put(10.5,-1){
\includegraphics[width=75mm,angle=0,keepaspectratio]{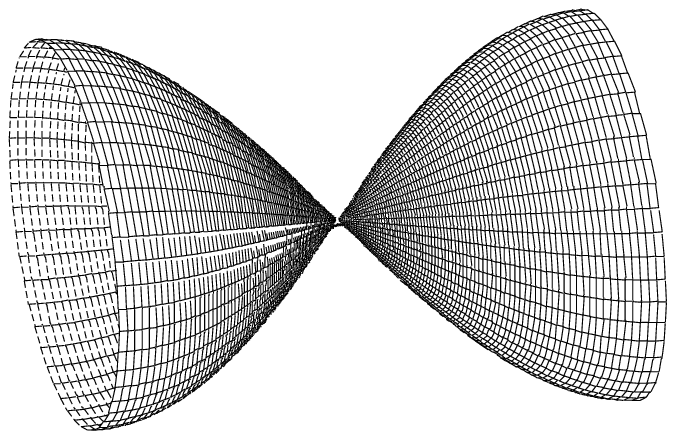}
}
%\vspace{-2cm}
\end{picture}
  \caption{
Spatial embedding of the horizon
of $D=5$ nonuniform black string solutions
approaching the horizon topology changing transition.
}
\label{f-2}
\end{figure}

Solutions with the new topology then correspond to nonuniform
black strings (NUBS), i.e., black strings whose horizon size
is not constant w.r.t.~the compact direction, but depends
on the compact coordinate \cite{Gubser:2001ac,Wiseman:2002zc}.
Examples of such nonuniform black strings with
increasing nonuniformity are shown
in Fig.~\ref{f-2}.

The simplest vacuum static solution of this type, however,
possesses translational symmetry along the
extracoordinate direction, and corresponds to a uniform black string
with horizon topology $S^{D-3}\times S^1$.
Although this solution exists for all values of the mass,
it is unstable below a critical value
as shown by Gregory and Laflamme \cite{Gregory:1993vy}.
Precisely at the marginally stable solution
the branch of NUBS arises.

Here we discuss the branches of static NUBS in $D=5$ and $D=6$
dimensions as well as the branches of rotating NUBS in $D=6$ dimensions,
where we again choose both angular momenta to be of equal-magnitude.
In all cases we present evidence for horizon topology
changing transitions.

\section{Rotating Einstein-Maxwell Black Holes}

\subsection{Generalities}

$D$-dimensional Einstein-Maxwell theory is based on
the Lagrangian
\begin{equation}
L = \frac{1}{16 \pi G_D} \sqdetg  (R - F_{\mu \nu} F^{\mu \nu}) \ ,
\end{equation}
with curvature scalar $R$,
$D$-dimensional Newton constant $G_D$,
and field strength tensor
$
F_{\mu \nu} =
\partial_\mu A_\nu -\partial_\nu A_\mu $,
where $A_\mu $ denotes the gauge potential.

Variation of the action with respect to the metric and the gauge potential
leads to the Einstein equations
\begin{equation}
G_{\mu\nu}= R_{\mu\nu}-\frac{1}{2}g_{\mu\nu}R = 2 T_{\mu\nu}
\ , \label{ee}
\end{equation}
with stress-energy tensor
\begin{equation}
T_{\mu \nu} = F_{\mu\rho} {F_\nu}^\rho - \frac{1}{4} g_{\mu \nu} F_{\rho
  \sigma} F^{\rho \sigma} \ ,
\end{equation}
and the gauge field equations,
\begin{equation}
\nabla_\mu F^{\mu\nu}  = 0 \ .
\label{feqA}
\end{equation}

We consider stationary black hole space-times with 
$N=[(D-1)/2]$ azimuthal symmetries,
implying the existence of $N+1$ commuting Killing vectors,
$\xi \equiv \partial_t$,
and $\eta_i \equiv \partial_{\vphi_i}$, for $i=1, \dots , N$.
The (constant) horizon angular velocities $\Omega_i$ 
can be defined by imposing the Killing vector field
\begin{equation}
\chi = \xi + \sum_{i=1}^N \Omega_i \eta_i \  \label{chi}
\end{equation}
to be null on and orthogonal to the horizon,
located at $r_{\rm H}$.
The horizon electrostatic potential $\Phi_{\rm H}$,
\begin{equation}
\Phi_{\rm H} =
\left. \chi^\mu A_\mu \right|_{r_{\rm H}} 
\ , \label{phiH} \end{equation}
is constant on the horizon, 
and likewise the surface gravity $\kappa$,
\begin{equation}
\kappa^2 = -\frac{1}{2} \left. (\nabla_\mu \chi_\nu) (\nabla^\mu \chi^\nu) 
\right|_{r_{\rm H}} \ .
\end{equation}

Asymptotically flat EM black holes 
with spherical horizon topology $S^{D-2}$
are then characterized by their mass $M$, charge $Q$, and
$N$ angular momenta $J_i$. 
The mass $M$ and the angular momenta $J_i$ of the black holes 
are obtained from the Komar expressions 
associated with the respective Killing vector fields
\begin{equation}
M = \frac{-1}{16 \pi G_D} \frac{D-2}{D-3} \int_{S_{\infty}^{D-2}} \alpha \ , \ \ \
J_i = \frac{1}{16 \pi G_D}  \int_{S_{\infty}^{D-2}} \beta_{(i)} \ ,
\end{equation}
with $\alpha_{\mu_1 \dots \mu_{D-2}} \equiv \epsilon_{\mu_1 \dots \mu_{D-2}
  \rho \sigma} \nabla^\rho \xi^\sigma$,
$\beta_{ (i) \mu_1 \dots \mu_{D-2}} \equiv \epsilon_{\mu_1 \dots \mu_{D-2}
  \rho \sigma} \nabla^\rho \eta_i^\sigma$.

The electric charge is obtained from
\begin{equation}
Q=\frac{-1}{8 \pi G_D} \int_{S_{\infty}^{D-2}} \tilde F \ ,
\end{equation}
with ${\tilde F}_{\mu_1 \dots \mu_{D-2}} \equiv  \epsilon_{\mu_1 \dots \mu_{D-2} \rho \sigma} F^{\rho \sigma}$.

The horizon mass $M_{\rm H}$ and horizon angular momenta
$J_{{\rm H},i}$ are given by
\begin{equation}
M_{\rm H} = \frac{-1}{16 \pi G_D} \frac{D-2}{D-3} \int_{{\cal H}} \alpha \ , \ \ \ 
J_{{\rm H},i} = \frac{1}{16 \pi G_D}  \int_{{\cal H}} \beta_{(i)} \ ,
\end{equation}
where ${\cal H}$ represents the surface of the horizon.

These black holes satisfy the first law \cite{Gauntlett:1998fz}
\begin{equation}
\rd M= \frac{\kappa}{8\pi G_D} \rd {A_{\rm H}} + \Phi_{\rm H} \rd Q +
\sum_{i=1}^N {\Omega_i} \rd {J_i}\, 
\label{first}
\end{equation}
and the generalized Smarr formula
\begin{equation}
M=  \frac{(D-2)}{(D-3)8\pi G_D} \kappa {A_{\rm H}}
+ \Phi_{\rm H} Q + 
\sum_{i=1}^N \frac{(D-2)}{(D-3)}{\Omega}_{i} {J_i} \ .
\label{smarr}
\end{equation}

\subsection{Metric and Gauge Potential}

We now exploit the enhanced symmetry of the black hole solutions
in odd dimensions,
arising when all $N$ angular momenta have equal-magnitude,
and parametrize the metric in isotropic coordinates,
which are well suited for the numerical construction of
rotating black holes 
\cite{Kunz:2005nm,Kunz:2006eh,Kunz:2005ei,Kunz:2006yp,Kunz:2006xk,Kleihaus:2000kg,Kleihaus:2002ee,Kleihaus:2002tc}.

The metric for these black holes with equal-magnitude angular momenta then
reads \cite{Kunz:2006eh}
\begin{eqnarray}
&&ds^2 = -f dt^2 + \frac{m}{f} \left[ dr^2 + r^2 \sum_{i=1}^{N-1}
  \left(\prod_{j=0}^{i-1} \cos^2\theta_j \right) d\theta_i^2\right] \nonumber \\
&&+\frac{n}{f} r^2 \sum_{k=1}^N \left( \prod_{l=0}^{k-1} \cos^2 \theta_l
  \right) \sin^2\theta_k \left(\vepsilon_k d\vphi_k - \frac{\omega}{r}
  dt\right)^2 \nonumber \\
&&+\frac{m-n}{f} r^2 \left\{ \sum_{k=1}^N \left( \prod_{l=0}^{k-1} \cos^2
  \theta_l \right) \sin^2\theta_k  d\vphi_k^2 \right. \nonumber\\
&& -\left. \left[\sum_{k=1}^N \left( \prod_{l=0}^{k-1} \cos^2
  \theta_l \right) \sin^2\theta_k \vepsilon_k d\vphi_k\right]^2 \right\} \ ,
\end{eqnarray}
where $\theta_0 \equiv 0$, $\theta_i \in [0,\pi/2]$
for $i=1,\dots , N-1$,
$\theta_N \equiv \pi/2$, $\vphi_k \in [0,2\pi]$ for $k=1,\dots , N$,
and $\vepsilon_k = \pm 1$ denotes the sense of rotation
in the $k$-th orthogonal plane of rotation.

An adequate parametrization for the gauge potential is given by
\begin{equation}
A_\mu dx^\mu =  a_0 dt + a_\vphi \sum_{k=1}^N \left(\prod_{l=0}^{k-1}
  \cos^2\theta_l\right) \sin^2\theta_k \vepsilon_k d\vphi_k \ .
\end{equation}
Thus, independent of the odd dimension $D\ge 5$,
this parametrization involves only four functions $f, m, n, \omega$
for the metric and two functions $a_0, a_\vphi$
for the gauge field, which all depend only on the radial coordinate $r$.

To obtain asymptotically flat solutions,
the metric functions should satisfy
at infinity the boundary conditions
\begin{equation}
f|_{r=\infty}=m|_{r=\infty}=n|_{r=\infty}=1 \ , \ \omega|_{r=\infty}=0 \ ,
\label{bc1} \end{equation}
while for the gauge potential we choose a gauge, in which it vanishes
at infinity
\begin{equation}
a_0|_{r=\infty}=a_\vphi|_{r=\infty}=0 \ .
\label{bc2} \end{equation}

The horizon is characterized by the condition $f(r_{\rm H})=0$ 
\cite{Kleihaus:2000kg}.
Requiring the horizon to be regular, the metric functions must
satisfy the boundary conditions
\begin{equation}
f|_{r=r_{\rm H}}=m|_{r=r_{\rm H}}=n|_{r=r_{\rm H}}=0 \ ,
\ \omega|_{r=r_{\rm H}}=r_{\rm H} \Omega \ ,
\label{bc3} \end{equation}
where $\Omega=\pm \Omega_i$ is the (up to a sign single constant) 
horizon angular velocity.
The gauge potential satisfies
\begin{equation}
\Phi_{\rm H} = \left. (a_0+\Omega a_\vphi)\right|_{r=r_{\rm H}} \ , \ \ \
\left. \frac{d a_\vphi}{d r}\right|_{r=r_{\rm H}}=0
\ . \label{bc4} \end{equation}

For equal-magnitude angular momenta $J=\pm J_i$, 
$i=1, \dots , N$, and likewise $J_{\rm H}= \pm J_{{\rm H},i}$.
The gyromagnetic ratio $g$ is then defined via
\begin{equation}
{\mu_{\rm mag}}=g \frac{Q J}{2M}
\ . \end{equation}
The global charges and the magnetic moment $\mu_{\rm mag}$
can be obtained from the asympotic expansions of the metric and the gauge
potential.

The enhanced symmetry can also be exploited to
obtain numerically charged rotating black holes 
in the presence of a cosmological constant \cite{Kunz:2007jq,Brihaye:2007bi}.

\subsection{Results}

\begin{figure}
\includegraphics[width=70mm,angle=0,keepaspectratio]{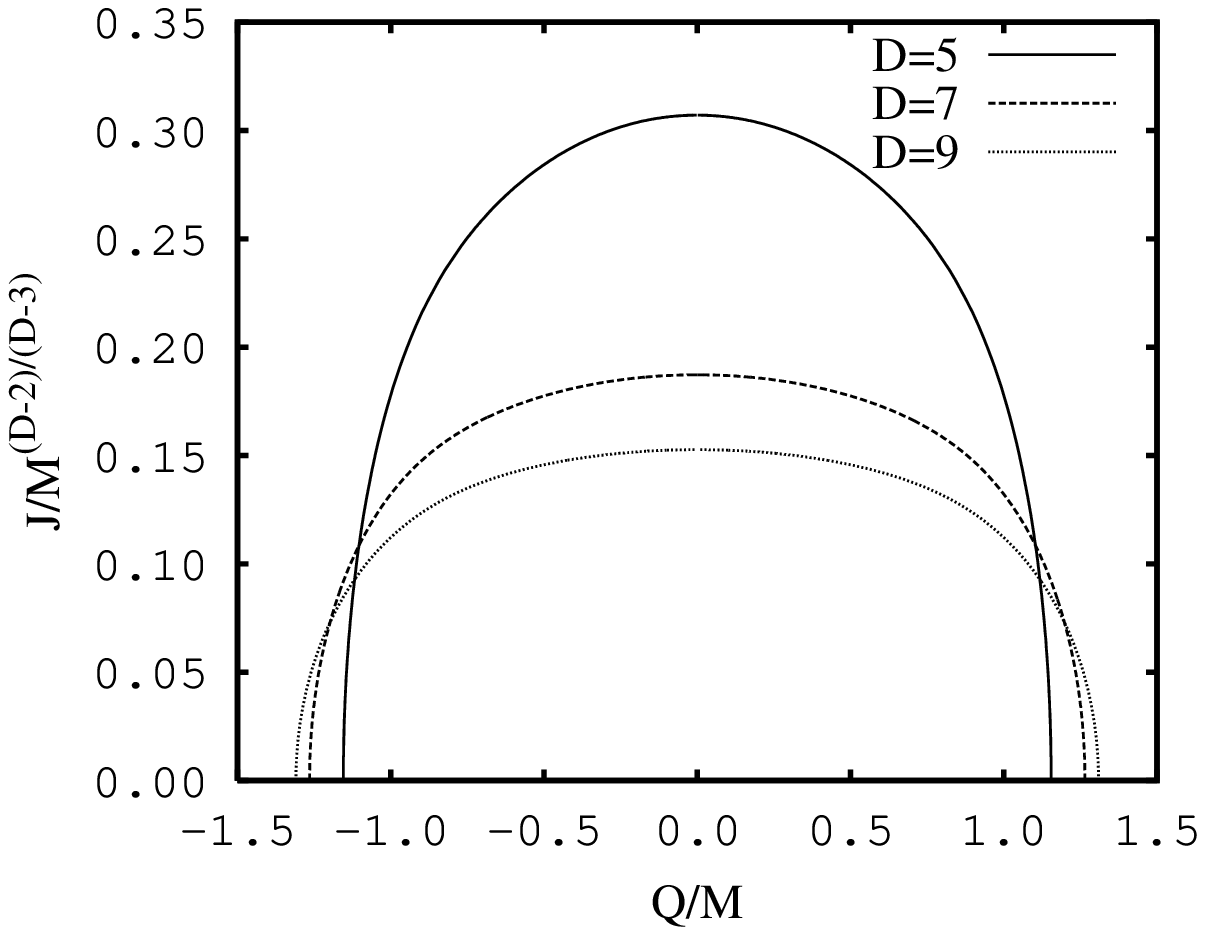}
\includegraphics[width=75mm,angle=0,keepaspectratio]{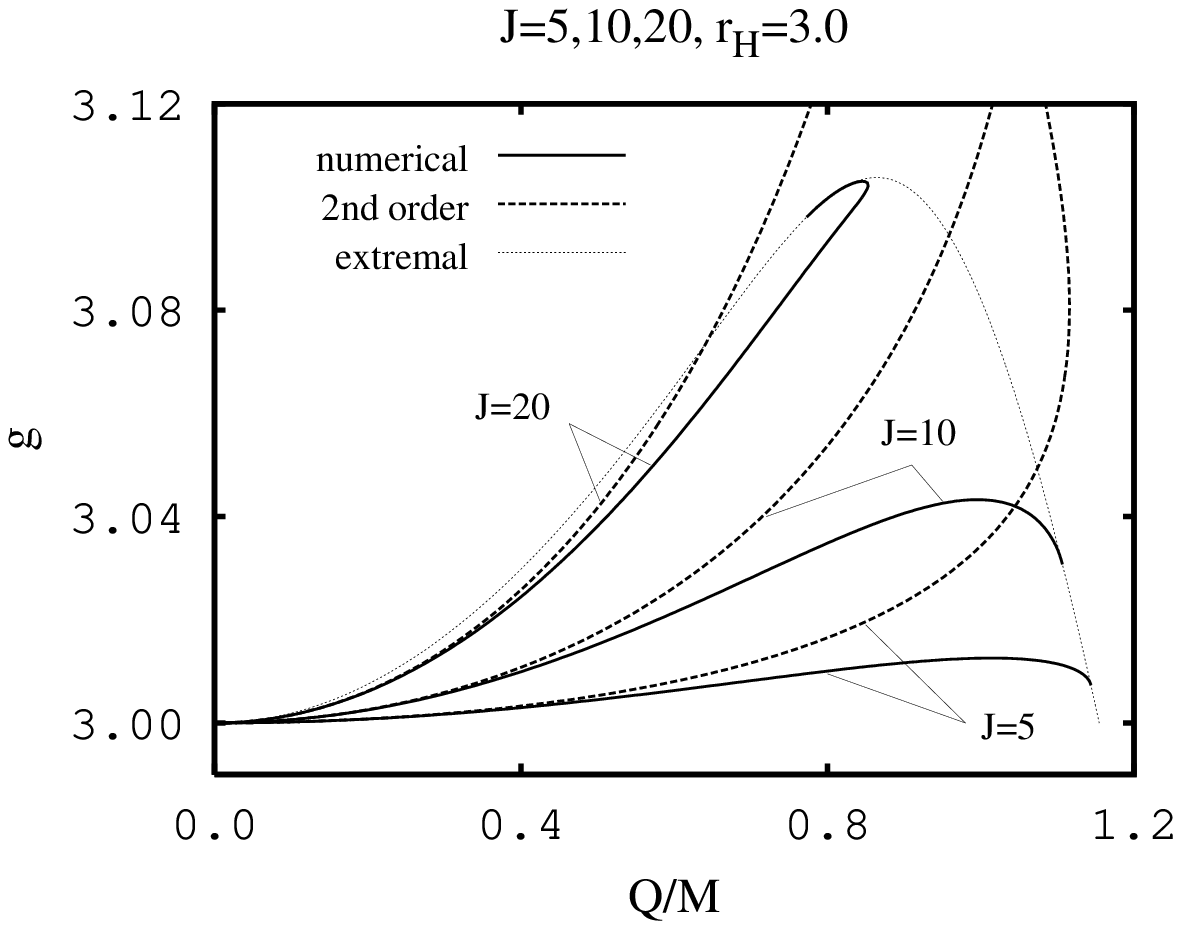}
  \caption{Left: Scaled angular momentum $J/M^{(D-2)/(D-3)}$ versus
scaled charge $Q/M$ for extremal black holes
with equal-magnitude angular momenta
in $D=5$, 7 and 9 dimensions.
Right: Gyromagnetic ratio $g$ versus the scaled electric charge $Q/M$ for
non-extremal black holes with horizon radius $r_{\rm H}=3.0$
and angular momentum $J=5$, 10 and 20:
numerical (solid), 2nd order perturbation (dashed)
(for comparison: $g$ for extremal solutions (thin-dotted)).}
\label{f-3}
\end{figure}

Let us first address the domain of existence
of rotating EM black holes with equal-magnitude
angular momenta.
We note, that unlike the case of a single non-vanishing angular momentum,
where no extremal solutions exist in $D>5$ dimensions 
\cite{Myers:1986un,Horowitz:1995tm},
extremal solutions always exist for odd $D$ black holes with
equal-magnitude angular momenta.
We exhibit in Fig.~\ref{f-3}
the scaled angular momentum $j=J/M^{(D-2)/(D-3)}$
of the extremal EM black holes
versus the scaled charge $q=Q/M$ % \cite{foot1}
for $D=5$, 7 and 9 dimensions \cite{Kunz:2006eh}.
Black holes exist only in the regions bounded by the
$J=0$-axis and by the respective curves.
The domain of existence is symmetric with respect to $Q \rightarrow -Q$.
These extremal black holes have vanishing surface gravity,
but finite horizon area.

The gyromagnetic ratio of Kerr-Newman black holes has the
constant value $g=2$ of Dirac particles.
In accordance with this value,
the gyromagnetic ratio has been obtained perturbatively
for rotating black holes in $D$ dimensions, yielding 
for small values of the charge $Q$
the constant lowest-order perturbative value $g=D-2$ 
\cite{Aliev:2004ec,Aliev:2006yk}.
However, second-order perturbative calculations 
reveal a non-constant value for the gyromagnetic ratio,
with, in particular, a quadratic dependence on the charge 
\cite{NavarroLerida:2007ez}.
We exhibit the gyromagnetic ratio in Fig.~\ref{f-3}
for extremal solutions, comparing these second-order results
with the full numerical results.
Thus the deviation of the gyromagnetic ratio
from the lowest-order perturbative value $g=D-2$ is always small,
but it is a true physical effect for higher-dimensional EM black holes.

\section{Einstein-Maxwell-Chern-Simons Black Holes}

\subsection{Supersymmetric black holes}

The extremal limits of the $D=5$ rotating charged black hole
solutions of EMCS theory (\ref{Lag}) with $\lambda=1$
are of special interest, since they encompass a
two parameter family of stationary supersymmetric black holes 
\cite{Breckenridge:1996is}.
The mass of these supersymmetric black holes
is given in terms of their charge
and saturates the bound \cite{Gibbons:1993xt}
\begin{equation}
M \ge \frac{\sqrt{3}}{2} |Q| \ ,
\label{M-bound}
\end{equation}
and their two equal-magnitude angular momenta, $|J|=|J_1|=|J_2|$,
are finite and satisfy the bound 
\cite{Gauntlett:1998fz,Herdeiro:2000ap,Townsend:2002yf}
\begin{equation}
%\frac{|J|}{M^{3/2}} \le \frac{1}{2} \left( \frac{\sqrt{3} }{2}
% \frac{Q}{M} \right)^{3/2}  \le \frac{1}{2} \ .
|J| \le \frac{1}{2} \left( \frac{\sqrt{3} }{2} {|Q|} \right)^{3/2} \, .
%J^2/M^3 \le 1/2 \ .
\label{J-bound}
\end{equation}

Thus the mass of these solutions does not change, 
as angular momentum is added.
Concerning the first law this implies that
the horizon angular velocity $\Omega$ must vanish for these black holes.
Thus their horizon is non-rotating, although their angular momentum
is nonzero,
i.e., the staticity theorem does not generalize to these
higher-dimensional black holes.

As the total angular momentum is increased from its static limiting
value $J=0$,
angular momentum is built up in the Maxwell field behind and outside
the horizon.
In particular, a negative fraction of the total angular momentum
is stored in the Maxwell field behind the horizon \cite{Gauntlett:1998fz}.
Thus, while one expects frame dragging effects
to cause the horizon to rotate,
these effects are precisely counterbalanced by frame dragging effects,
due to the negative contribution to the angular momentum
by the fields behind the horizon,
allowing these black holes to retain a static horizon \cite{Townsend:2002yf}.

All these supersymmetric black holes possess
a regular horizon, except for the limiting solution,
saturating the bound Eq.~(\ref{J-bound}).
The area $A_{\rm H}$ of the horizon decreases as $|J|$ increases towards
the bound Eq.~(\ref{J-bound}),
yielding a singular limiting solution
with vanishing horizon area, $A_{\rm H}=0$.
The effect of the rotation on the horizon is not to make it rotate,
but to deform it from a round 3-sphere to a squashed 3-sphere
\cite{Gauntlett:1998fz}.

These special properties of $D=5$ supersymmetric EMCS black holes
caused intriguing speculations on how the properties
of $D=5$ black holes in general EMCS theories depend
on the CS coefficient \cite{Gauntlett:1998fz}.
In particular, 
since the supersymmetric black holes appear to be marginally stable,
these speculations predicted instability for extremal EMCS black holes,
when the CS coefficient would be increased beyond its
supergravity value $\lambda = 1$.

The argument is as follows \cite{Gauntlett:1998fz}:
Extremal static EMCS black holes
saturate the bound for any value of $\lambda$.
If the mass of extremal stationary black holes
decreases with increasing $\lambda$ for fixed angular momentum,
and increases with increasing angular momentum for fixed $\lambda<1$,
while it is independent of angular momentum for $\lambda=1$,
it becomes possible that the mass can
decrease with increasing angular momentum for fixed $\lambda>1$.

Thus while an extremal static black hole
with zero Hawking temperature and spherical symmetry
cannot decrease its mass by Hawking radiation,
it can however become unstable with respect to rotation, when $\lambda>1$,
with photons carrying away both energy and angular momentum
to infinity.
In terms of the first law as applied to $\lambda>1$
extremal black holes ($\kappa=0$) with fixed charge ($\rd Q=0$),
such an instability then requires, that
the horizon is rotating in the opposite sense to the angular momentum,
since $\rd M = 2 {\Omega} \rd {J}$
has to be negative.

\boldmath
\subsection{EMCS black holes: $\lambda>1$}
\unboldmath

\begin{figure}[h!]
\includegraphics[width=70mm,angle=0,keepaspectratio]{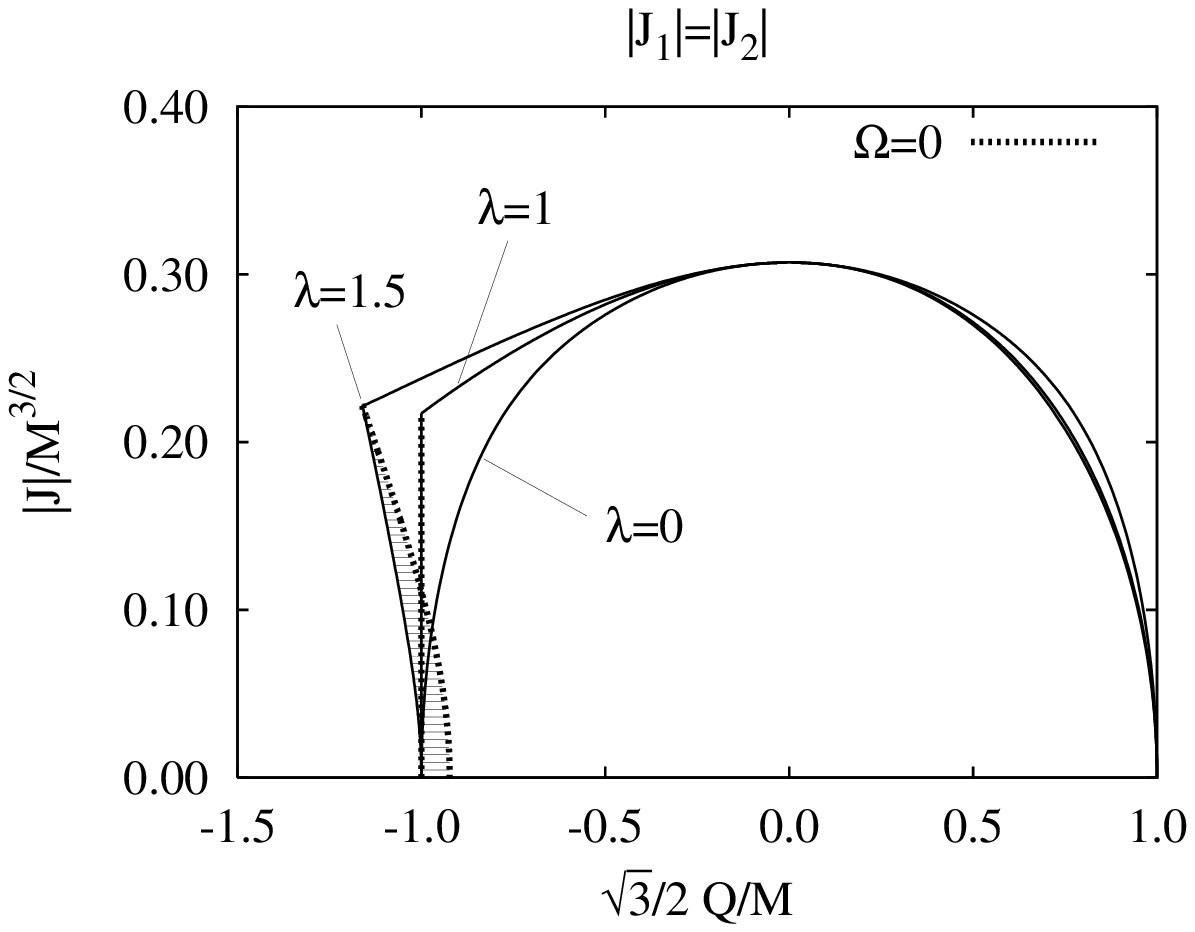}
\includegraphics[width=70mm,angle=0,keepaspectratio]{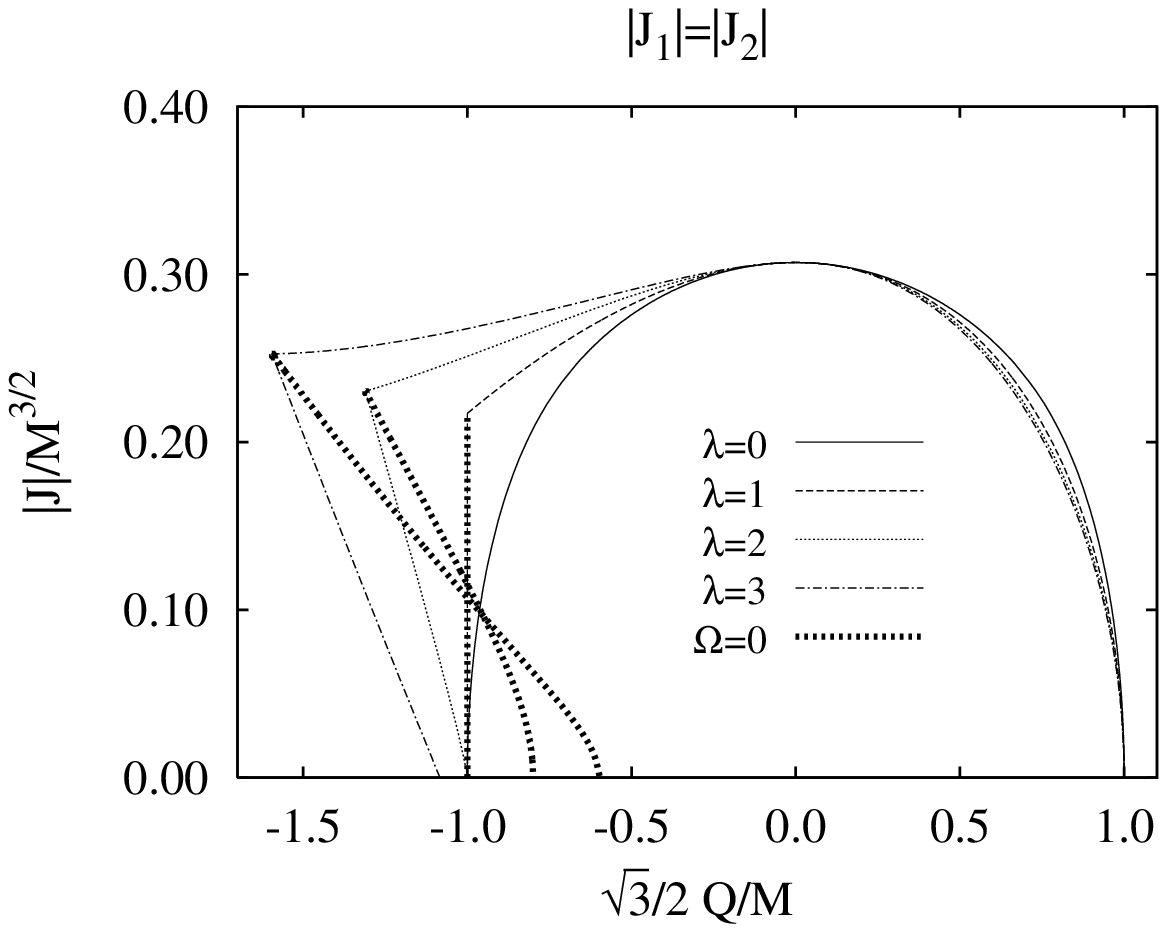}
  \caption{
Scaled angular momentum $J/M^{3/2}$ versus
scaled charge $Q/M$ for extremal black holes (outer curves)
and stationary black holes with non-rotating horizon (dotted curves)
(left: $\lambda=0$, 1, 1.5), (right: $\lambda=0$, 1, 2, 3).
}
\label{f-4}
\end{figure}

Our numerical results for $D=5$ EMCS black holes with equal-magnitude
angular momenta confirm these predictions.
A summary of some of the main results is exhibited in Fig.~\ref{f-4}.
Here the scaled angular momentum $|J|/M^{3/2}$ of the extremal EMCS black holes
is shown versus the scaled charge $Q/M$
for several values of $\lambda$:
the pure EM case, $\lambda=0$ \cite{Kunz:2005nm},
the supergravity case, $\lambda=1$ \cite{Breckenridge:1996is},
and $\lambda=1.5$, $\lambda=2$, 
and $\lambda=3$ \cite{Kunz:2005ei}.
For a given value of $\lambda$,
black holes exist only in the regions bounded by the
$J=0$-axis and by the respective outer curves.
(Note the asymmetry of domain of existence of the black hole solutions
with respect to $Q \rightarrow -Q$
for non-vanishing CS term.)

\begin{figure}[h!]
\includegraphics[width=70mm,angle=0,keepaspectratio]{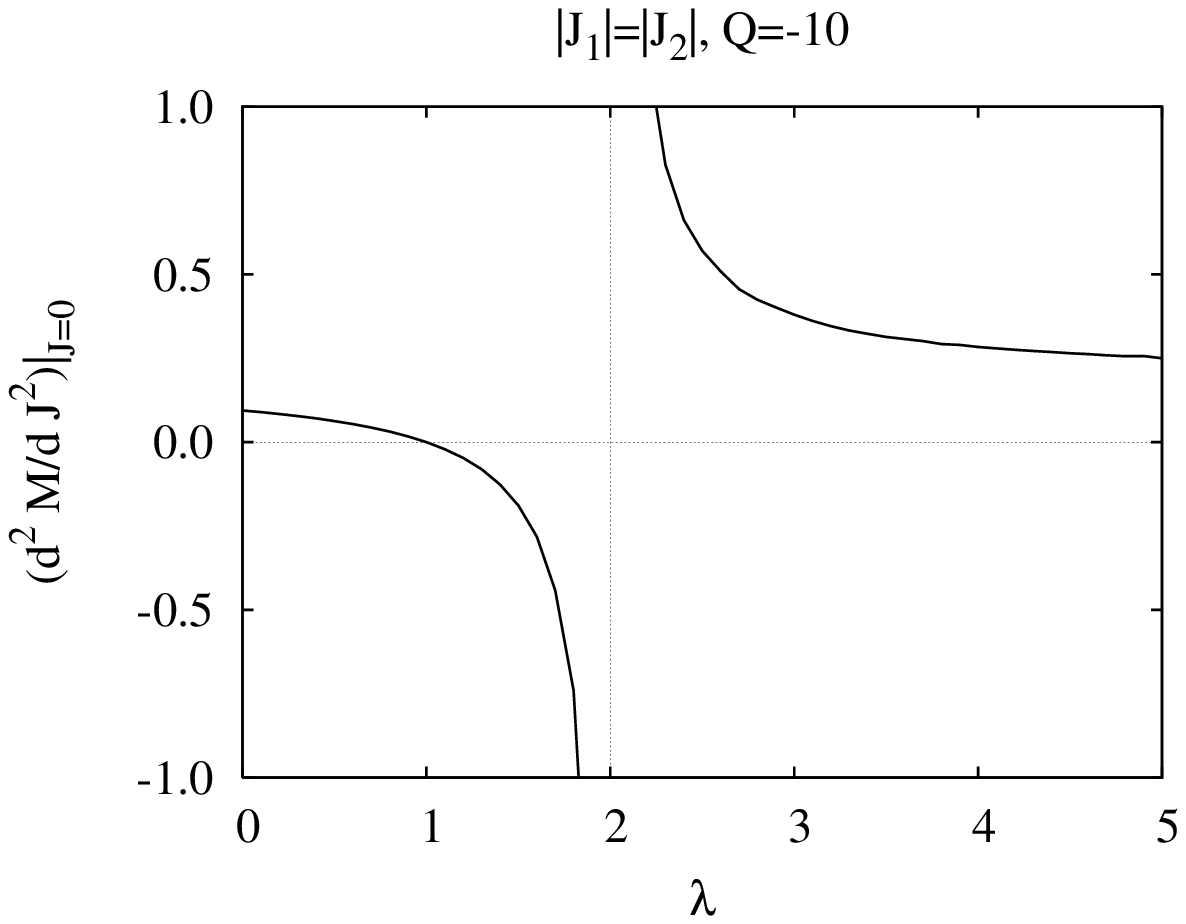}
\includegraphics[width=70mm,angle=0,keepaspectratio]{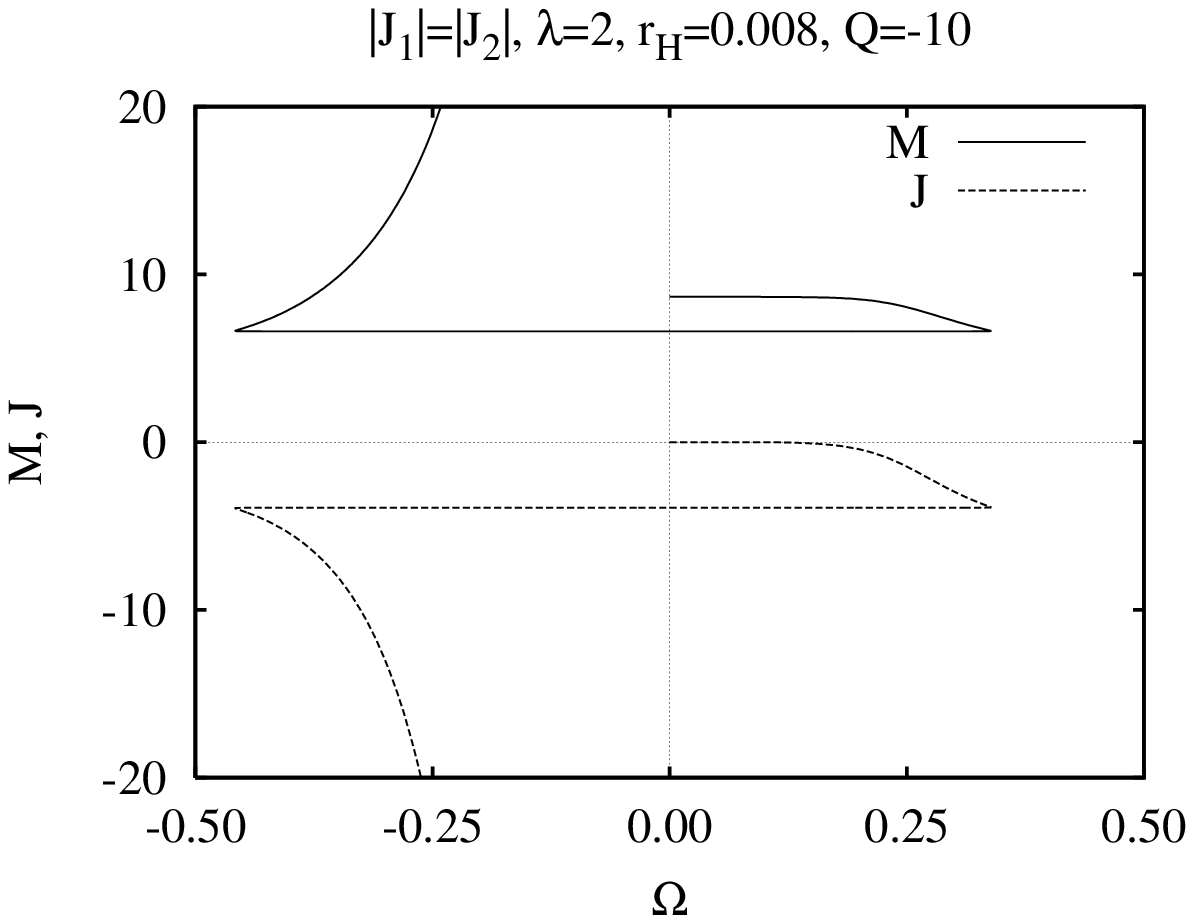}
  \caption{
Left: Second-order derivative of the mass $M$ with respect to the angular momentum
$J$ at $J=0$ versus the CS coupling constant $\lambda$ for extremal
black holes ($Q=-10$).
Right: Angular momentum $J$ and mass $M$
versus horizon angular velocity $\Omega$ for almost extremal black holes
($\lambda=2$, $r_{\rm H}=0.008$, $Q=-10$).
\label{f-5}
}
\end{figure}

In Fig.~\ref{f-5} we demonstrate explicitly,
that extremal static black holes become
unstable with respect to rotation beyond $\lambda=1$:
For $1< \lambda < 2$ the mass decreases
with increasing magnitude of the angular momentum for fixed electric charge,
since counterrotating solutions arise here.
Thus supersymmetry marks the
borderline between stability and instability for these black hole solutions.

For a given value of $\lambda \le 2$,
the set of rotating $\Omega=0$ black holes
then divides the domain of existence into two parts.
The right part contains ordinary black holes,
where the horizon rotates in the same sense as the angular momentum,
whereas the left part 
contains counterrotating black holes,
i.e., their horizon rotates in the opposite sense to the angular momentum
\cite{Kunz:2005ei,Kunz:2006yp,Kleihaus:2003df}.

\begin{figure}[h!]
\includegraphics[width=70mm,angle=0,keepaspectratio]{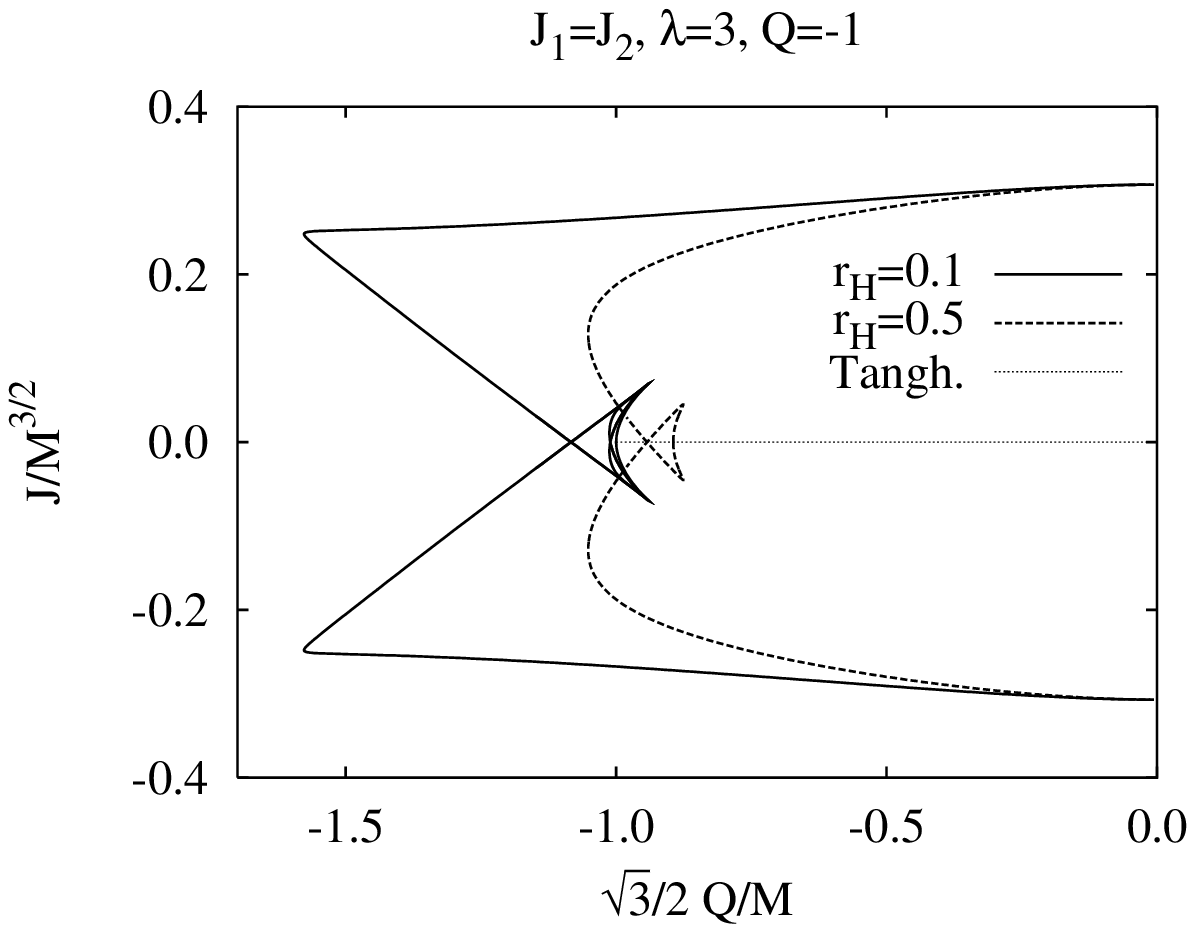}
\includegraphics[width=70mm,angle=0,keepaspectratio]{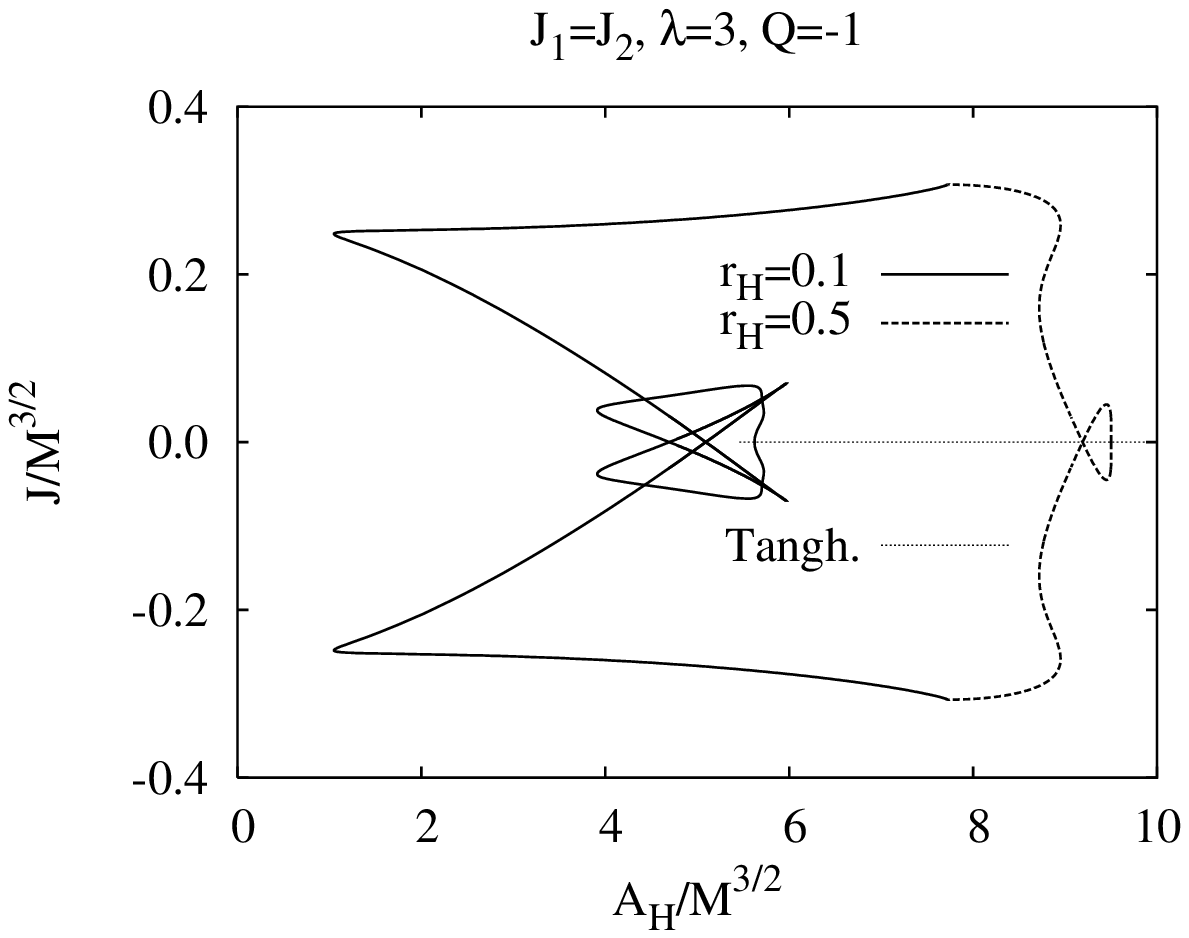}
  \caption{
Left: Scaled angular momentum $J/M^{3/2}$ versus scaled charge $Q/M$
(left) and versus scaled horizon volume $A_{\rm H}/M^{3/2}$ (right)
for non-extremal black holes with horizon radii $r_{\rm H}=0.1$ and $0.5$
($\lambda=3$; $Q=-1$).
}
\label{f-6}
\end{figure}

\begin{figure}[h!]
\setlength{\unitlength}{1cm}
\begin{picture}(15,12)
%\put(-1,0){\epsfig{file=Fig2l0.5new.eps,width=8cm}}
\put(0,0){
\includegraphics[width=70mm,angle=0,keepaspectratio]{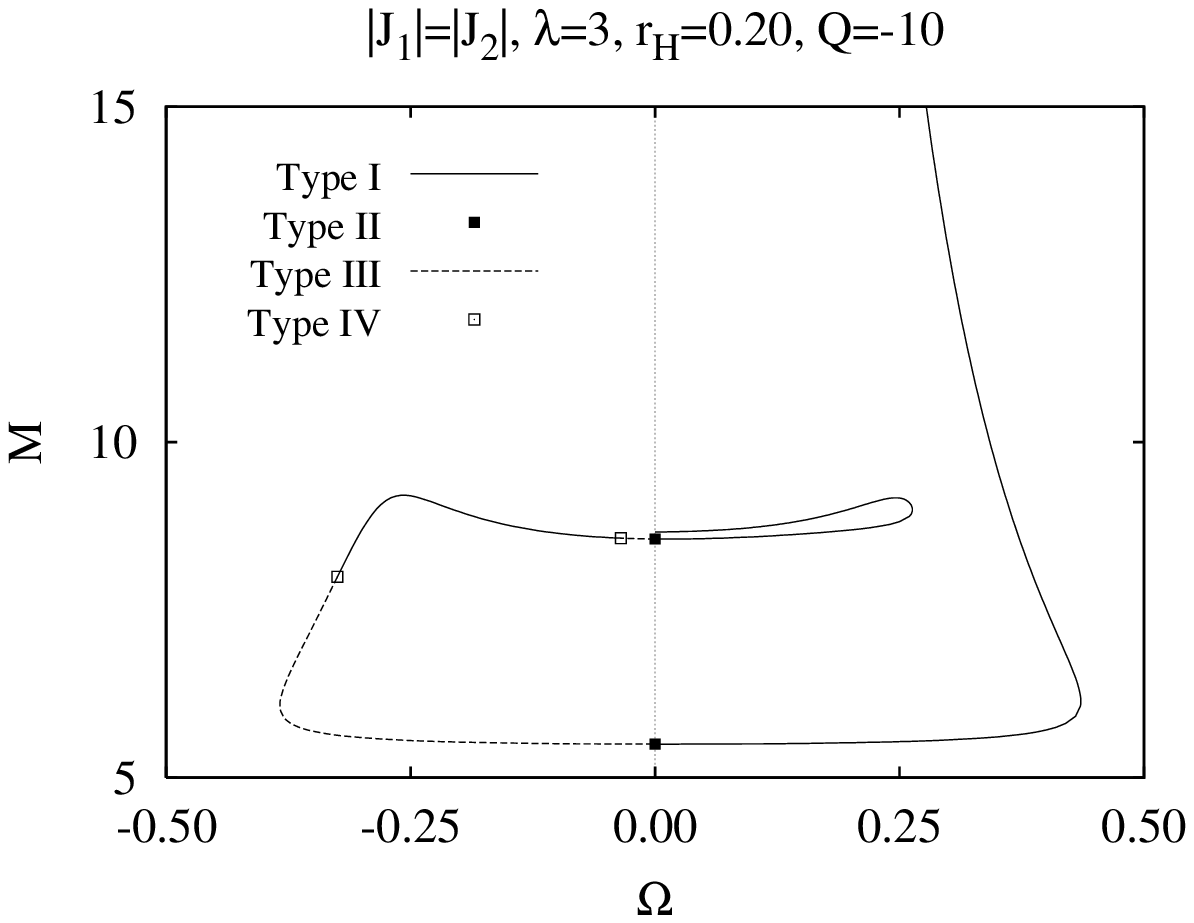}
}
\put(7,0){
\includegraphics[width=70mm,angle=0,keepaspectratio]{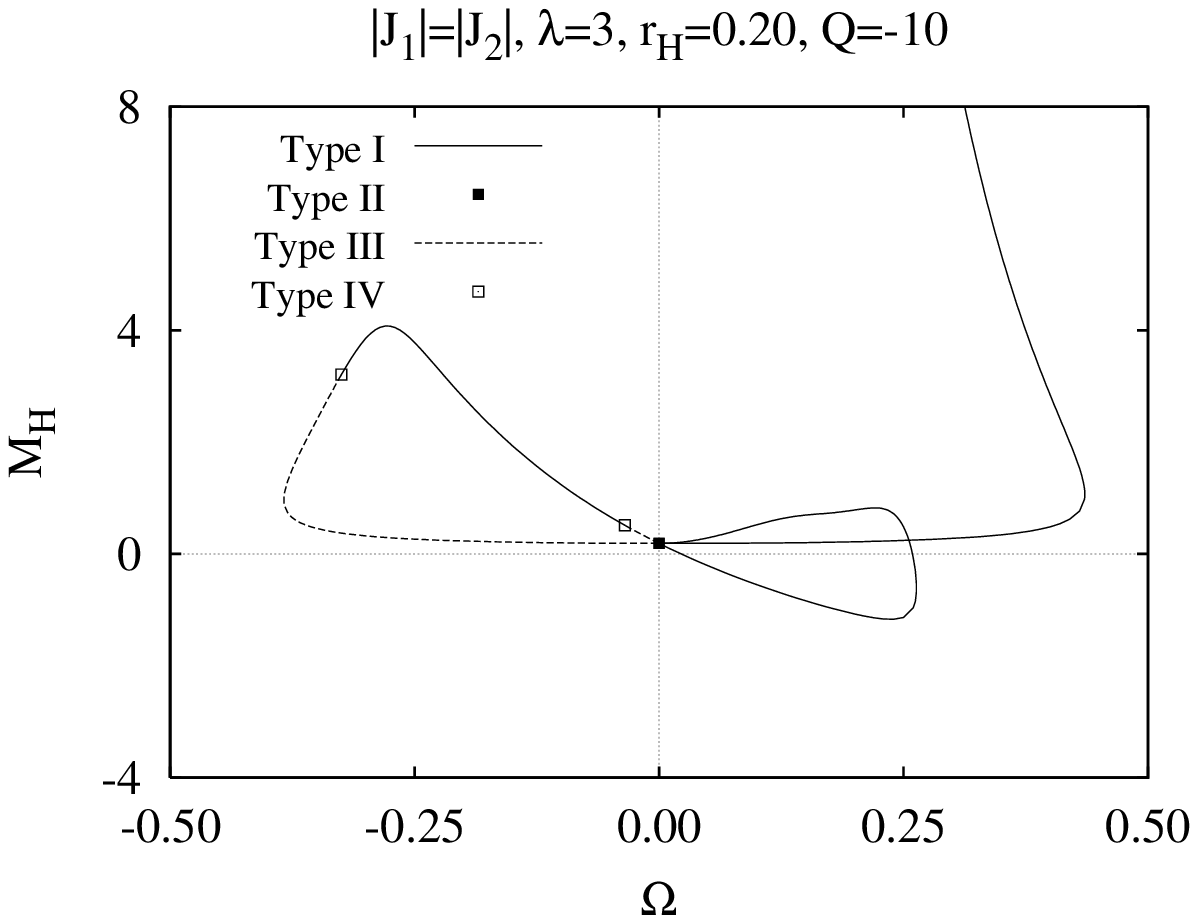}
}
\put(0,6){
\includegraphics[width=70mm,angle=0,keepaspectratio]{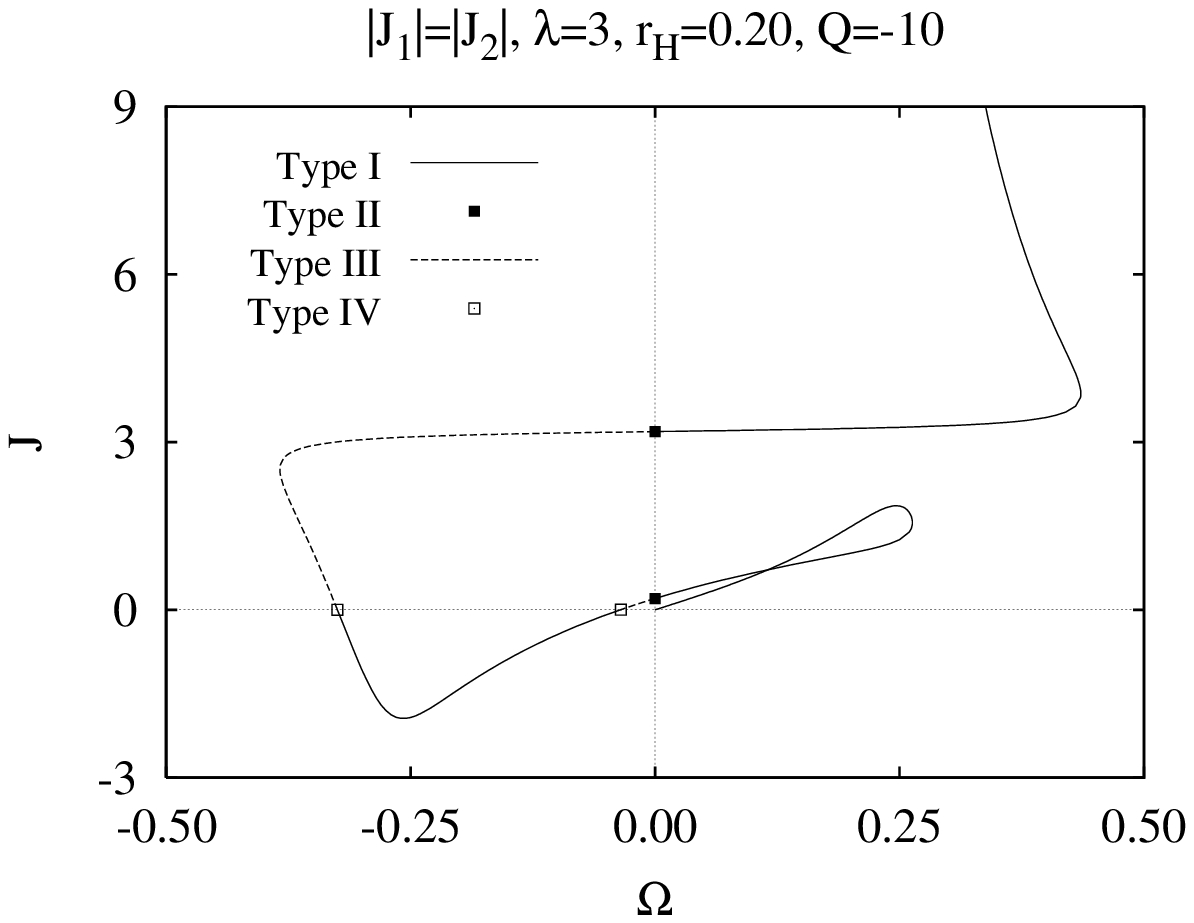}
}
\put(7,6){
\includegraphics[width=70mm,angle=0,keepaspectratio]{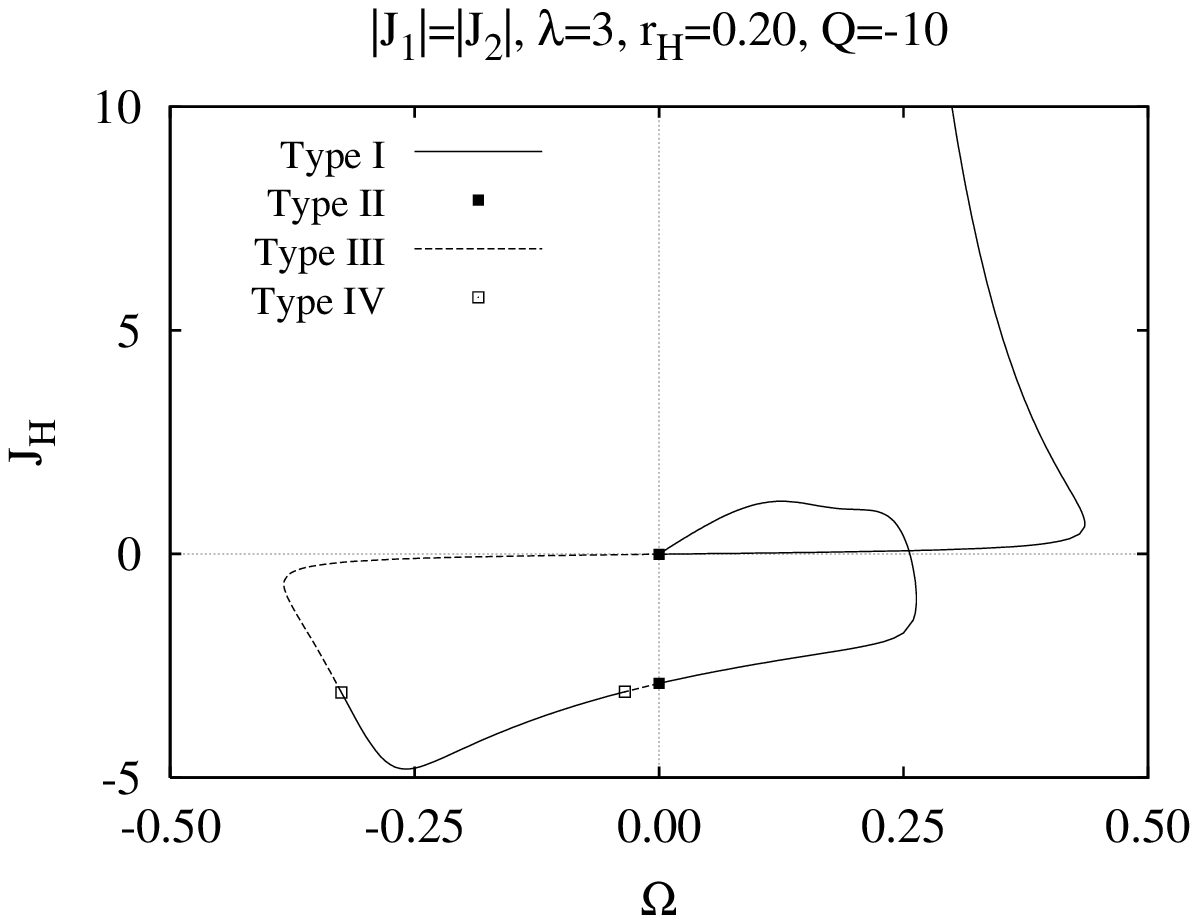}
}
\end{picture}
\caption{
Properties of non-extremal $\lambda=3$ EMCS black holes with
charge $Q=-10$ and horizon radius $r_{\rm H}=0.2$.
Angular momentum $J$ (upper left), 
horizon angular momentum $J_{\rm H}$ (upper right),
mass $M$ (lower left), horizon mass $M_{\rm H}$ (lower right)
versus horizon angular velocity $\Omega$.}
\label{f-7}
\end{figure}

As expected from the change in stability,
another special case is reached, when $\lambda=2$.
Indeed, for $\lambda=2$
the numerical analysis indicates (see Fig.~\ref{f-5}), that
a (continuous) set of rotating $J=0$ solutions appears
and persists as $\lambda$ is increased beyond the value $\lambda=2$
\cite{Brodbeck:1997ek}.
The existence of these rotating $J=0$ solutions
relies on a special partition of the total
angular momentum $J$, where the angular momentum within the horizon
$J_{\rm H}$ is precisely equal and opposite to
the angular momentum in the Maxwell field outside the horizon.
In contrast, for $\lambda<2$ only static $J=0$ solutions exist.
The presence of $J=0$ solutions is also seen in Fig.~\ref{f-6}
and Fig.~\ref{f-7} for $\lambda=3$.

Fig.~\ref{f-6} further reveals that beyond $\lambda=2$
black holes are no longer uniquely characterized by their global charges.
Thus the uniqueness theorem does not generalize to $D=5$ EMCS
stationary black holes with horizons of spherical topology,
provided $\lambda>2$.
The previous counterexamples to black hole uniqueness involved black rings,
i.e., black objects with the horizon topology of a torus
\cite{Emparan:2001wn}.
For $\lambda=2$ even an infinite set of extremal
black holes with the same global charges appears to exist,
as numerical data indicate.

To explore the properties of $\lambda>2$ EMCS black holes further,
let us now consider non-extremal black holes.
We exhibit in Fig.~\ref{f-7} a set of solutions for $\lambda=3$,
possessing constant charge $Q=-10$ and
constant (isotropic) horizon radius $r_{\rm H}=0.2$.
In particular, we exhibit the total angular momentum $J$
the horizon angular momentum $J_{\rm H}$, 
the mass $M$ and the horizon mass $M_{\rm H}$
versus the horizon angular velocity $\Omega$.

There are four types of rotating black holes,
as classified by their total angular momentum $J$
and horizon angular velocity $\Omega$: Type I black
holes correspond to the corotating regime, i.e., $\Omega J \ge 0$, and
$\Omega=0$ if and only if $J=0$ (static). 
Type II black holes possess a static horizon
($\Omega=0$), although their angular momentum does not vanish ($J\neq 0$).
Type III black holes are characterized by counterrotation,
i.e., the horizon angular velocity
and the total angular momentum have oposite signs, $\Omega J < 0$.
Type IV black holes, finally, possess a rotating horizon ($\Omega \neq
0$) but vanishing total angular momentum ($J=0$).

As the horizon of the black hole is set into rotation,
angular momentum is stored in the Maxwell field both behind
and outside the horizon, yielding a rich variety of configurations.
In particular, even when the solutions are corotating,
i.e., $J$ and $\Omega$ rotate in the same sense,
$J_{\rm H}$ and $\Omega$ can assume opposite signs.
Then the product $\Omega J_{\rm H}$ turns negative,
and can give rise to black holes
with negative horizon mass, $M_{\rm H}<0$,
as seen in Fig.~\ref{f-7}.
Thus the negative fraction of the angular momentum 
stored in the Maxwell field behind the horizon
is responsible for the occurrence of a negative horizon mass.
The total mass is always positive, however.

\boldmath
\subsection{EMCS black holes: $D>5$}
\unboldmath

In higher odd dimensions the corresponding CS term
yields a modified Smarr formula,
which is supplemented by an additional term.
This term is proportional to the CS coefficient $\lambda$ and
to $(D-5)$ \cite{Gauntlett:1998fz}, i.e., $D=5$ is a rather special case
among the class of odd-dimensional
Einstein-Maxwell-Chern-Simons (EMCS) theories,
since the Smarr formula (\ref{smarr}) remains unmodified.

All the new types of stationary black holes
found in $5D$ EMCS theory
occur also for EMCS black holes
in higher odd dimensions \cite{Kunz:2006yp}.
But in higher dimensions
the CS coefficient becomes dimensionful
and changes under scaling transformations, unless
${\lambda=0}$.
Thus any feature present for a certain
non-vanishing value of ${\lambda}$
will be present for any other non-vanishing value
(although for the correspondingly scaled value of the charge).
Most interestingly, in $D=9$, a further type of 
stationary black holes appears:
V. non-static $\Omega=J=0$ black holes,
possessing a finite magnetic moment \cite{Kunz:2006yp}.

\section{Nonuniform Black Strings}

\subsection{Generalities}

Black string solutions approach asymptotically the
$D-1$ dimensional Minkowski-space times a circle ${\cal M}^{D-1}\times S^1$.
The coordinates of $R^{D-2}$ are denoted by $x^1,...,x^{D-2}$,
the compact coordinate by $z = x^{D-1}$, and $x^D=t$.
The radial coordinate $r$ is given by
$r^2 = (x^1)^2 + \cdots + (x^{D-2})^2$,
and the compact coordinate is periodic with period $L$.

Nonuniform black string solutions (NUBS) of the vacuum Einstein equations
can be obtained with the ansatz
\cite{Kleihaus:2006ee}
\begin{eqnarray}
\label{metric}
ds^2=-e^{2A(r,z)}f(r)dt^2+e^{2B(r,z)}\left(\frac{dr^2}{f(r)}+dz^2\right)+e^{2C(r,z)}r^2d\Omega_{D-3}^2
, \end{eqnarray}
where
\begin{eqnarray}
\nonumber
f=1-\left(\frac{r_{\rm H}}{r}\right)^{D-4}  .
\end{eqnarray}
The event horizon resides at a surface of constant radial coordinate
$r=r_{\rm H}$ and is characterized by the condition $f(r_{\rm H})=0$.

Utilizing the reflection symmetry of the NUBS
w.r.t.~$z=L/2$,
the solutions are subject to the following set of
boundary conditions
\begin{eqnarray}
\label{Nbc1}
A\big|_{r=\infty}=B\big|_{r=\infty}=C\big|_{r=\infty}=0,
\end{eqnarray}
\begin{eqnarray}
\label{Nbc2}
A\big|_{r=r_{\rm H}}-B\big|_{r=r_{\rm H}}=d_0,~\partial_{r}
A\big|_{r=r_{\rm H}}=\partial_{r} C\big|_{r=r_{\rm H}}=0,
\end{eqnarray}
\begin{eqnarray}
\label{Nbc3}
\partial_z A\big|_{z=0,L/2}=\partial_z B\big|_{z=0,L/2}
=\partial_z C\big|_{z=0,L/2}=0,
\end{eqnarray}
where the constant $d_0$ is related to the Hawking
temperature of the solutions.
Regularity further requires that the condition
$\partial_{r} B\big|_{r=r_{\rm H}}=0$ holds for the solutions.

For a static spacetime which is asymptotically
${\cal M}^{D-1}\times S^1$
one obtains two asymptotic quantities,
the mass $M$ and the tension ${\mathcal T}$ \cite{Traschen:2001pb}.
These are encoded in the asymptotics of the metric potentials.
With
\begin{eqnarray}
\label{1}
g_{tt}\simeq -1+\frac{c_t}{r^{D-4}},~~~g_{zz}\simeq 1+\frac{c_z}{r^{D-4}}
\, ,
\end{eqnarray}
the mass and tension of black string solutions are given by
\cite{Kol:2004ww,Harmark:2007md,Harmark:2003dg}
\begin{eqnarray}
\label{2}
M=\frac{\Omega_{D-3}L}{16 \pi G}((D-3)c_t-c_z),
~~{\mathcal T}=\frac{\Omega_{D-3}}{16 \pi G}(c_t-(D-3)c_z),
\end{eqnarray}
where $\Omega_{D-3}$ is the area of the unit $S^{D-3}$ sphere.
The corresponding quantities of the uniform black string (UBS) solutions
$M_0$ and ${\mathcal T}_0$ are obtained from (\ref{2}) with $c_z=0$,
$c_t=r_{\rm H}^{D-4}$.
The relative tension $n$,
\begin{eqnarray}
\label{3}
n=\frac{{\mathcal T} L}{M}=\frac{c_t-(D-3)c_z}{(D-3)c_t-c_z} \, ,
\end{eqnarray}
then measures how large the tension is relative to the mass.
This dimensionless quantity is bounded, $0\leq n\leq D-3$,
where the UBS have relative tension $n_0=1/(D-3)$.

For black strings the first law of thermodynamics reads
\begin{eqnarray}
\label{firstlaw}
dM=TdS+{\mathcal T}dL \, ,
\end{eqnarray}
and the Smarr formula can be expressed as
\begin{eqnarray}
\label{smarrform}
TS = \frac{D-3-n}{D-2} M \, .
\end{eqnarray}

\subsection{Static Nonuniform Black Strings}

The branch of nonuniform black strings emerges smoothly from
the uniform black string branch at the critical point,
where the stability of the UBS changes \cite{Gregory:1993vy}.
Keeping the horizon coordinate $r_{\rm H}$ and the asymptotic
length $L$ of the compact direction fixed,
the solutions depend on a single parameter, $d_0$,
specified via the boundary conditions.
Varying this parameter, the nonuniform strings
become increasingly deformed.
This nonuniformity is quantified by the
parameter $\Lambda$,
\begin{equation}
\Lambda = \frac{1}{2} \left( \frac{{\cal R}_{\rm max}}{{\cal R}_{\rm min}}
 -1 \right)
, \label{Lambda} \end{equation}
where ${\cal R}_{\rm max}$ and ${\cal R}_{\rm min}$
represent the maximum radius of a $(D-3)$-sphere on the horizon and
the minimum radius, being the radius of the `waist'.
Thus for uniform black strings $\Lambda=0$,
while the conjectured horizon topology
changing transition should be approached
for $\Lambda \rightarrow \infty$
\cite{Kol:2004ww,Harmark:2007md,Wiseman:2002ti,Kol:2003ja,Kudoh:2004hs}.

\begin{figure}[h!]
\setlength{\unitlength}{1cm}
\begin{picture}(15,18)
%\put(-1,0){\epsfig{file=Fig2l0.5new.eps,width=8cm}}
\put(-1,0){
\includegraphics[width=90mm,angle=0,keepaspectratio]{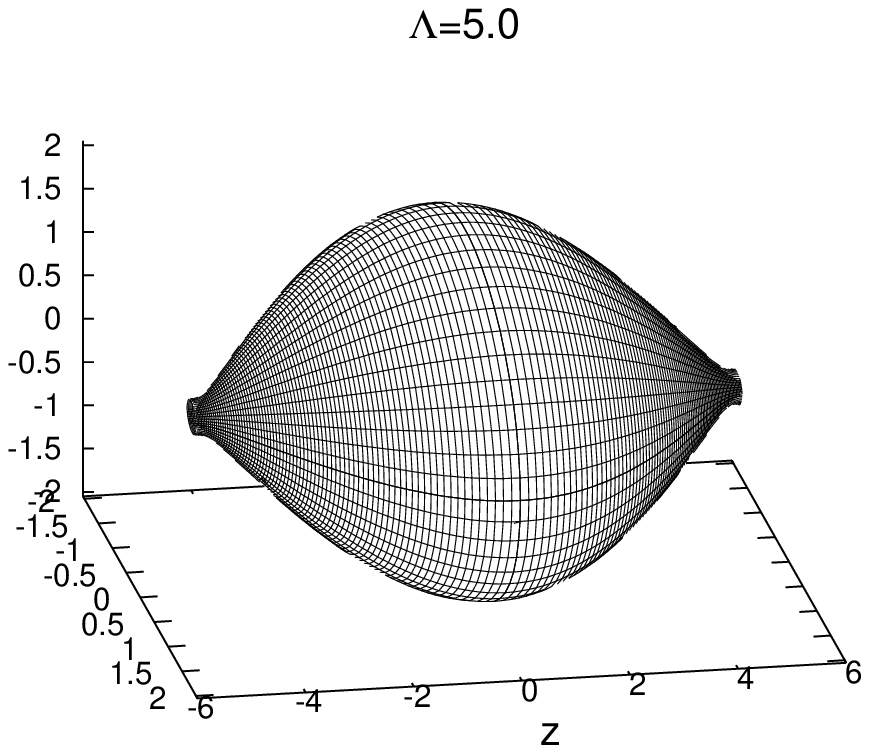}
}
\put(7,0){
\includegraphics[width=90mm,angle=0,keepaspectratio]{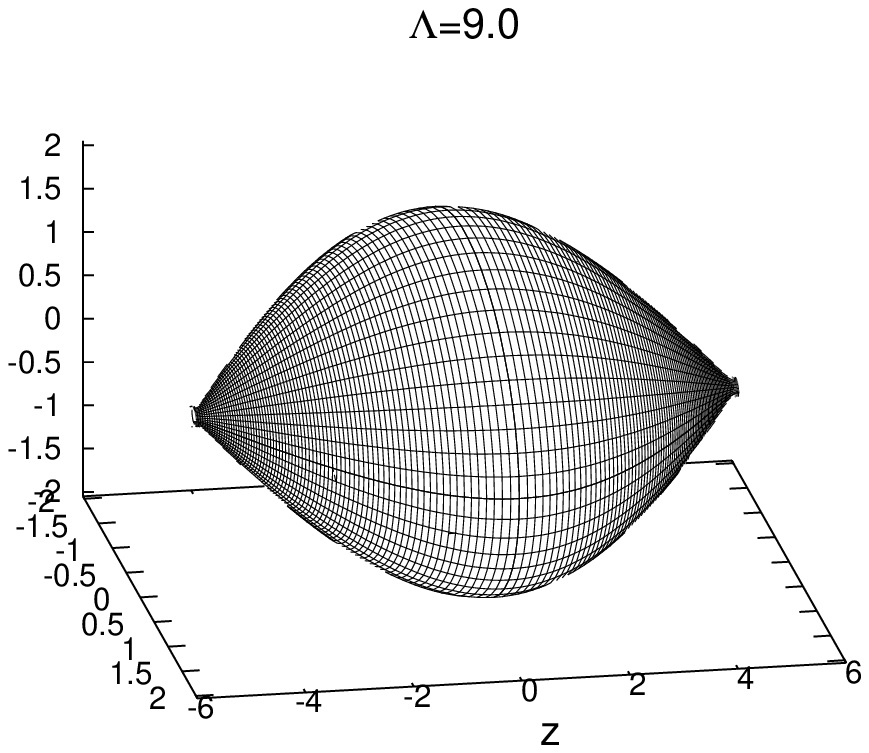}
}
\put(-1,6){
\includegraphics[width=90mm,angle=0,keepaspectratio]{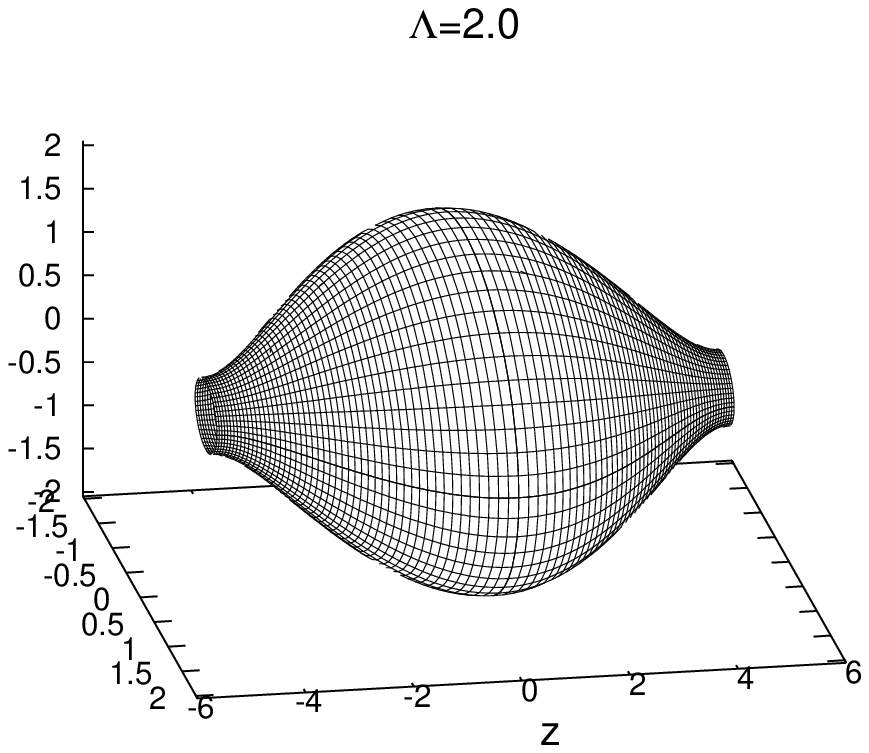}
}
\put(7,6){
\includegraphics[width=90mm,angle=0,keepaspectratio]{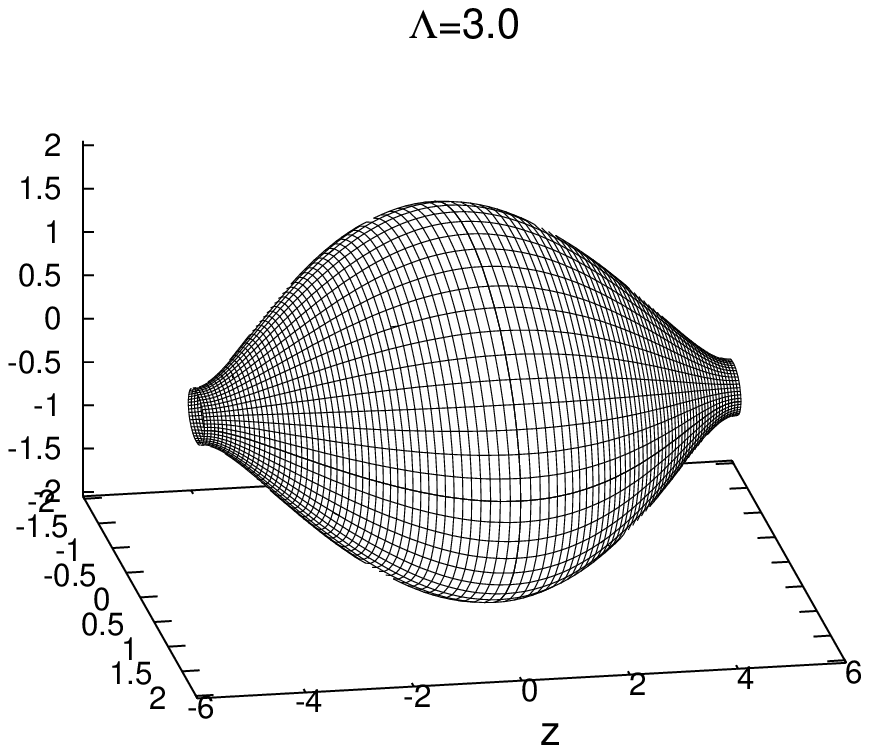}
}
\put(-1,12){
\includegraphics[width=90mm,angle=0,keepaspectratio]{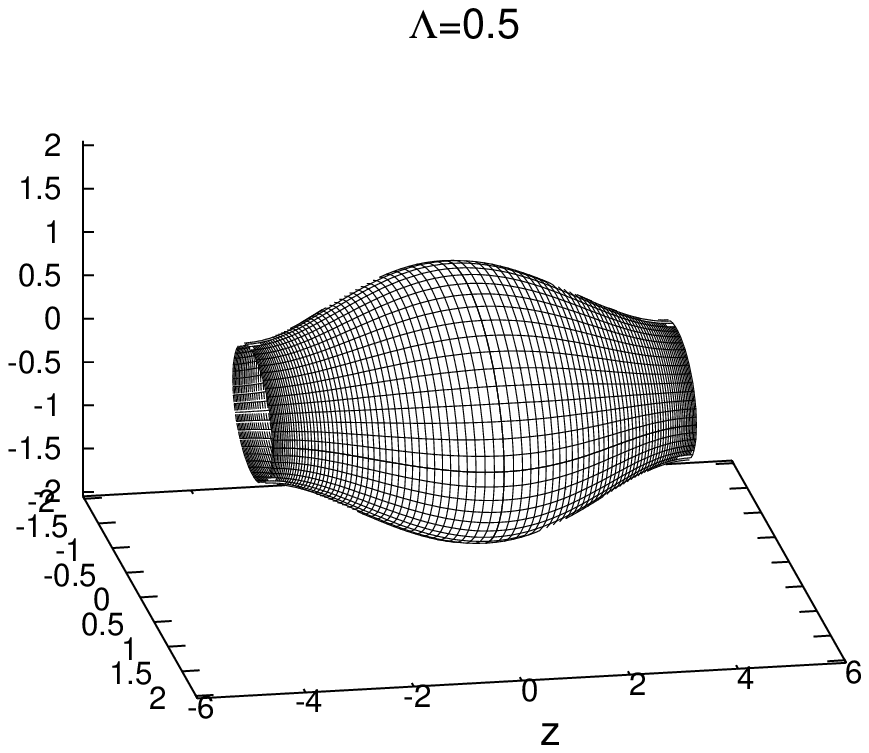}
}
\put(7,12){
\includegraphics[width=90mm,angle=0,keepaspectratio]{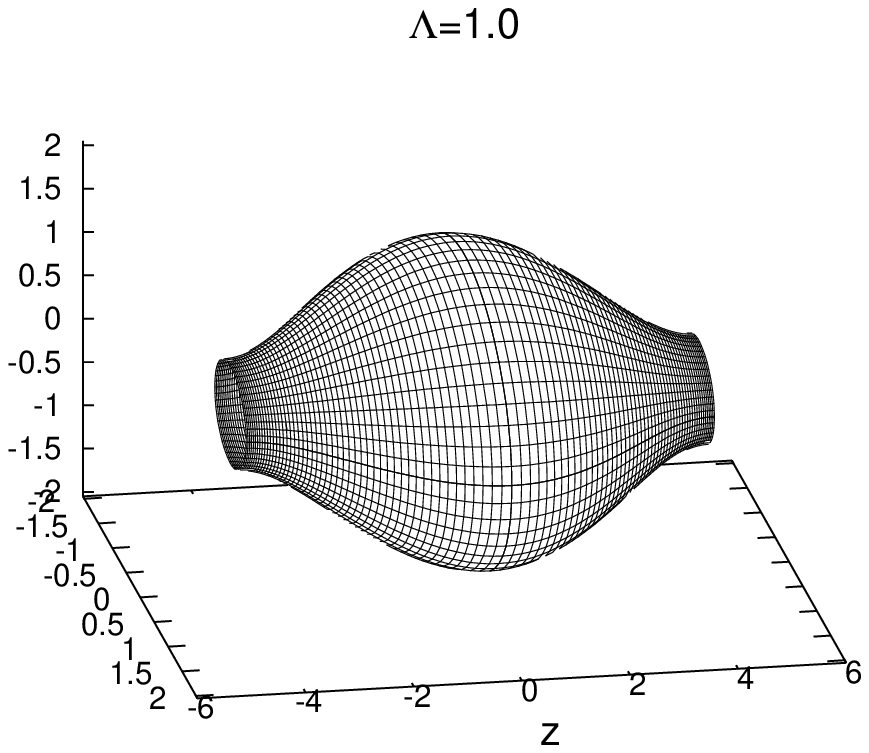}
}
\end{picture}
%{\small {\bf Figure 3.}
\caption{
The spatial embedding of the horizon
of $D=5$ nonuniform black string solutions
with horizon coordinate $r_{\rm H} =1$
and asymptotic length of the compact direction $L=L^{\rm crit}=7.1713$,
is shown for increasing nonuniformity, $\Lambda=0.5,~1,~2,~3,~5,~9$.
}
\label{f-8}
\end{figure}

We exhibit in Fig.~\ref{f-8}
the spatial embedding of the horizon into 3-dimensional space
for $D=5$ nonuniform black string solutions
with increasing nonuniformity \cite{Kleihaus:2006ee}.
In these embeddings the proper radius of the horizon is plotted
against the proper length along the compact direction,
yielding a geometrical view of the nonuniformity
of the solutions.
With increasing $\Lambda$, ${\cal R}_{\rm max}$ appears to
approach a finite value in the limit $\Lambda \rightarrow \infty$,
whereas ${\cal R}_{\rm min}$ appears to reach zero in this limit
(when extrapolated).

Evidence, that the nonuniform black string branch 
and the black hole branch merge at a
horizon topology changing transition,
was first provided in $D=6$ dimensions \cite{Kudoh:2004hs}:
By considering the mass, the entropy, and the temperature
of the branch of nonuniform black strings as well as of
the branch of caged black holes versus the relative tension $n$, 
it appeared likely, that both branches would merge
at a critical value $n_*$.

\begin{figure}[h!]
\includegraphics[width=70mm,angle=0,keepaspectratio]{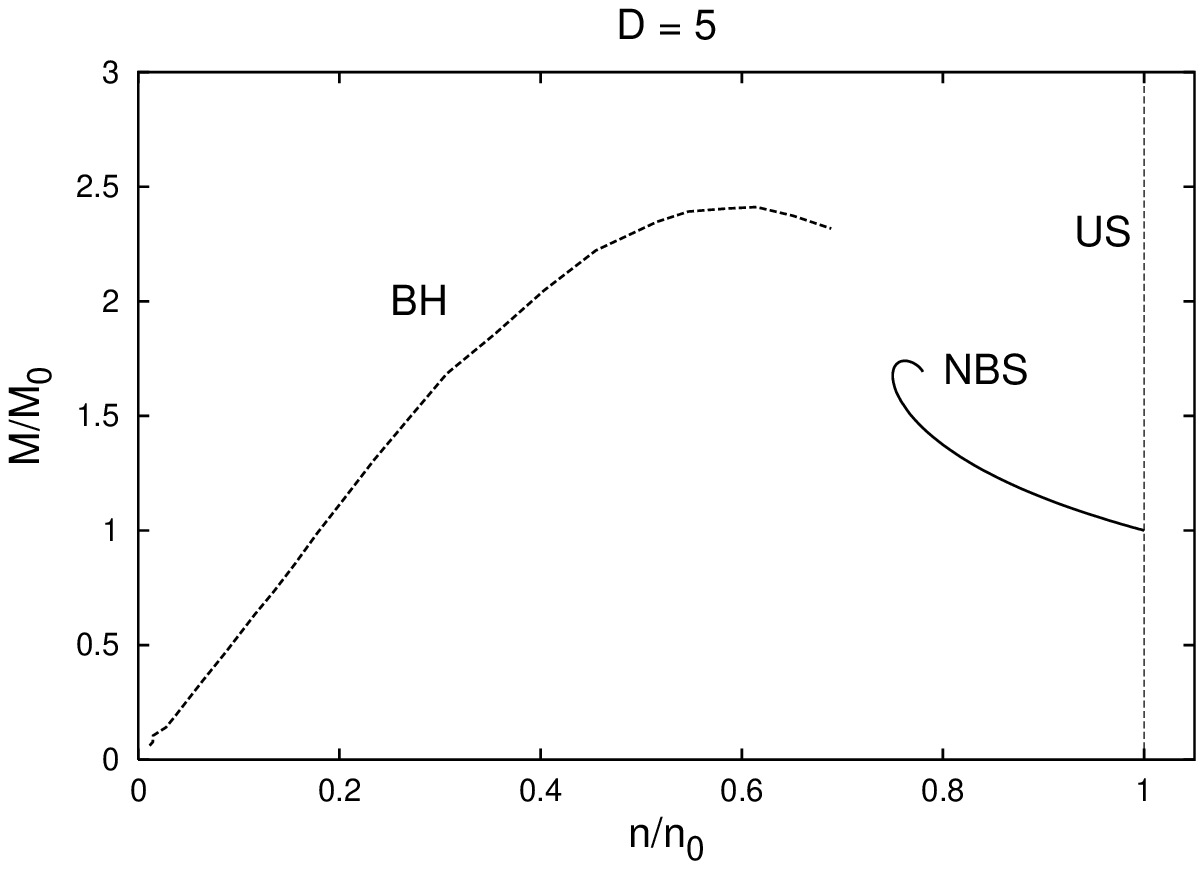}
\includegraphics[width=70mm,angle=0,keepaspectratio]{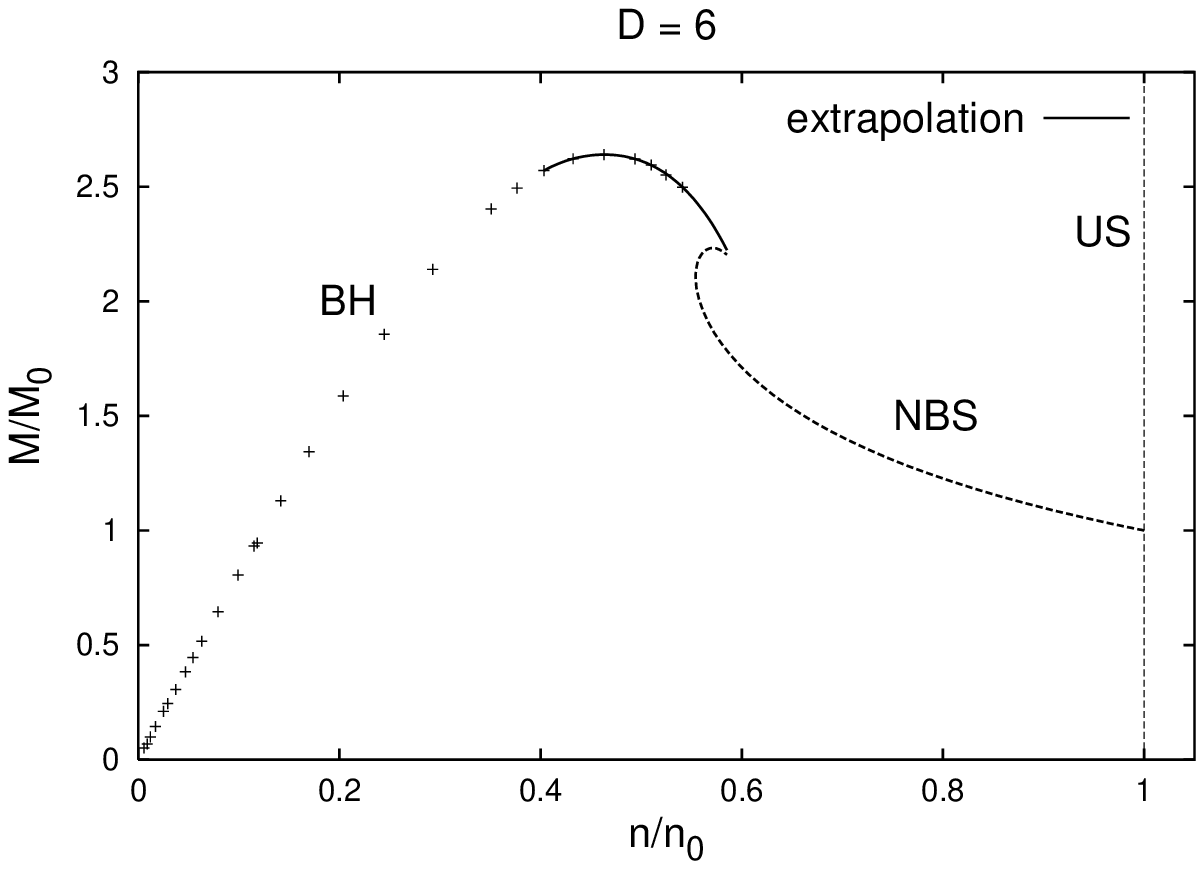}
  \caption{
The mass $M$ of the $D=5$ (left) and $D=6$ (right) nonuniform string
and black hole branches
versus the relative string tension $n$.
($M$ and $n$ are normalized by the values of the
corresponding uniform string solutions.)
The data for the black hole branches is from
\cite{Kudoh:2004hs}.
The $D=6$ black hole branch is extrapolated towards the
anticipated critical value $n_*$.
}
\label{f-9}
\end{figure}

Extending the nonuniform black string branch in $D=6$ dimensions
and obtaining for the first time 
the nonuniform black string branch in $D=5$ dimensions,
we observe a backbending of the nonuniform black string
branch in both $D=5$ and $D=6$ dimensions w.r.t.~$n$,
as shown in Fig.~\ref{f-9}.
Still, all our data are consistent with the assumption,
that the nonuniform string branch and the black hole branch
merge at a horizon topology changing transition.
In fact, extrapolation of the black hole branch towards
this transition point $n_*$ appears to
match well the (extrapolated) endpoint of the (backbending) part
of the nonuniform string branch (Fig.~\ref{f-9}).

For the phase diagram this means that we have a region
$0 < n < n_{\rm b}$ with one branch of black hole solutions, then a region
$n_{\rm b} < n < n_*$ with one branch of black hole solutions
and two branches of nonuniform string solutions, the ordinary one
and the backbending one,
and finally a region $n_* < n < n_0$ with only one branch of
nonuniform string solutions.
(We here do not address the bubble-black hole sequences present for $n > n_0$).
Thus the horizon topology changing transition is associated with $n_*$,
and $n_{\rm b} < n < n_*$ represents a middle region where
three phases coexist, one black hole and two nonuniform strings.
The anticipated phase diagram is seen in Fig.~\ref{f-9} (right).

This is strongly reminiscent of the phase structure of the
rotating black ring--rotating black hole system in $D=5$ \cite{Emparan:2001wn}.
The (asymptotically flat) rotating black holes have horizon topology $S^3$,
and the (asymptotically flat)
rotating black rings have horizon topology $S^2 \times S^1$.
The rotating black holes exist up to a maximal value of the
angular momentum (for a given mass), $0 < J < J_*$,
the rotating black rings are present only above a minimal value of the
angular momentum (for a given mass), $J_{\rm b} < J$,
and in the middle region $J_{\rm b} < J < J_*$ three phases coexist,
one black hole and two black rings \cite{Emparan:2001wn}.

Further numerical work for nonuniform strings and in particular
for black holes
in the critical region close to $n_*$
should confirm this anticipated phase diagram 
for nonuniform black strings and caged black holes
further and 
lead to further insight into the structure of the configuration space,
in particular in the region close to the horizon topology changing transition.

But not only further study of the nonuniform black string solutions
outside their horizon may be instructive, also study of their interior
should give insight into the expected
horizon topology changing transition.
We therefore now address the inside of nonuniform black string
solutions.

The interior solutions are obtained by solving the corresponding
set of hyperbolic equations, for which the values of the
respective functions at the horizon are known (from the elliptic
exterior problem) and employed as initial values.
As the integration proceeds inside the horizon, 
the singularity is encountered along the curve $r_{\rm s}(z)$
\cite{Kleihaus:2007cf}.

\begin{figure}[h!]
\includegraphics[width=70mm,angle=0,keepaspectratio]{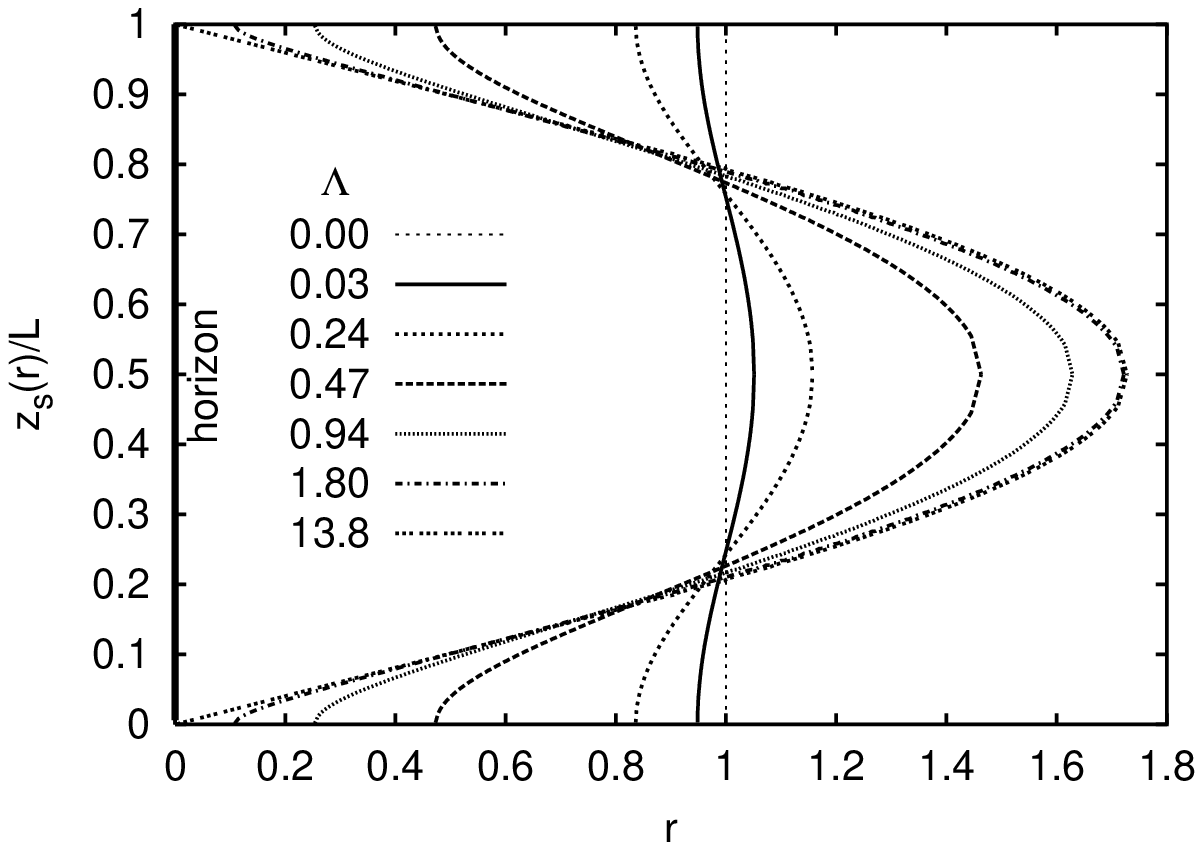}
\includegraphics[width=77mm,angle=0,keepaspectratio]{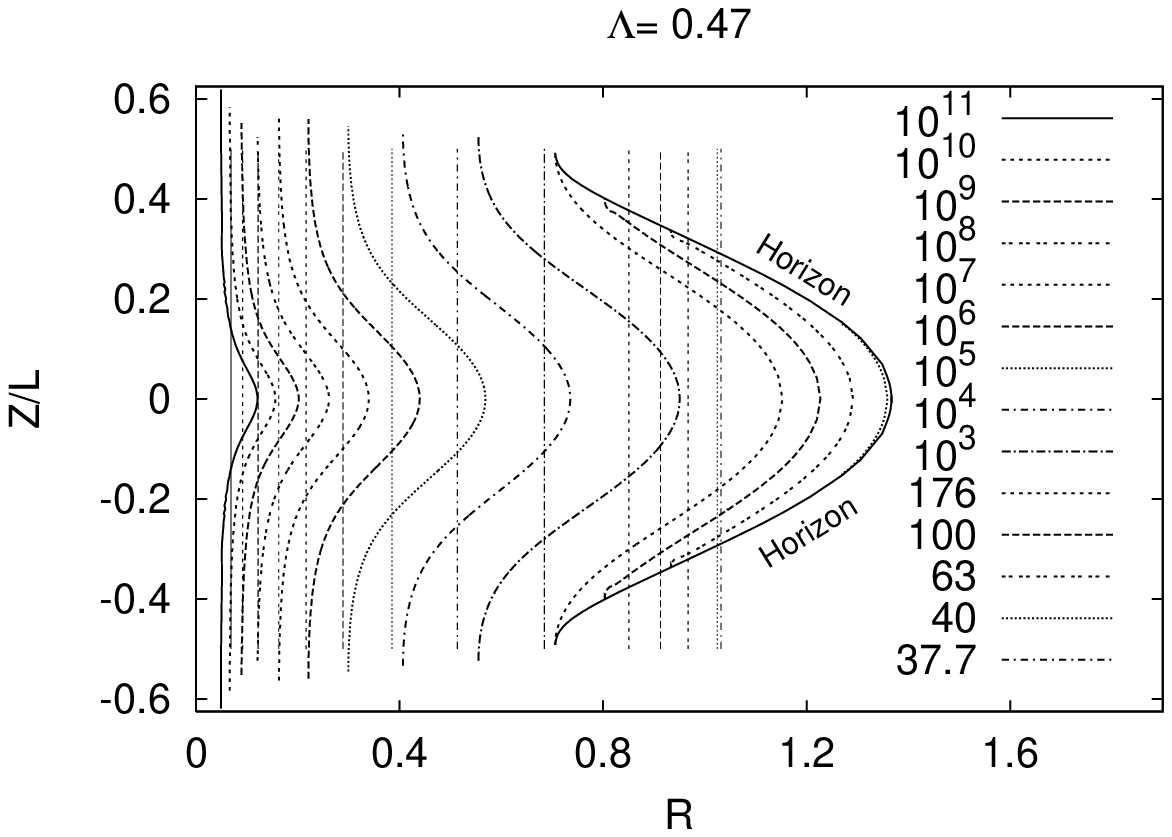}
  \caption{Left: 
The coordinates of the singularity $r_{\rm s}(z)$ for several
values of the nonuniformity parameter $\Lambda$.
(In these coordinates the horizon is located at $r=0$.)
Right:
Isometric embedding of surfaces of constant Kretschmann scalar
for the interior of a nonuniform black string with $\Lambda=0.47$.
The straight lines correspond to the uniform black string
with the same temperature.
}
\label{f-10}
\end{figure}

We exhibit the curve $r_{\rm s}(z)$ of the singularity
in Fig.~\ref{f-10} for nonuniform black string solutions
with increasing nonuniformity.
Note, that in these coordinates the horizon is located at $r=0$,
while the singularity of uniform black string solutions
resides at $r_{\rm s}=1$.
As the nonuniformity is turned on,
the coordinate of the singularity develops a periodic oscillation about
$r_{\rm s}=1$. 
Thus the singularity is located at $r_{\rm s}<1$ at the waist of the
nonuniform black string, and at $r_{\rm s}>1$ in the vicinity
of the maximal horizon radius of the string.
With increasing nonuniformity
this minimal value decreases monotonically 
while the maximal value increases monotonically.
When extrapolated, 
the minimal value then appears to touch the horizon in the limit,
where the horizon topology changing transition is 
expected to be reached.

In order to get more insight into the geometry of space
in the nonuniform black string interior,
we exhibit in Fig.~\ref{f-10} also isometric embeddings of
surfaces of constant Kretschmann scalar for the inside of a
nonuniform black string solution with
nonuniformity parameter $\Lambda=0.47$.

\subsection{Rotating Black Strings}

To obtain nonuniform generalizations of the rotating uniform black
string MP solutions,
we consider space-times with  $ (D-2)/2$ commuting Killing vectors
$ \partial_{\varphi_k}$.
While the general configuration will possess $ (D-2)/2 $ independent
angular momenta, we again restrict to rotating NUBS whose
angular momenta have all equal magnitude,
since analogous to the case of black holes \cite{Kunz:2006eh},
the metric parametrization then simplifies considerably
for such rotating NUBS.

In $D=6$ dimensions we have obtained numerically
such rotating nonuniform black strings with equal angular momenta
\cite{Kleihaus:2007dg}.
These emerge from the branch of
marginally stable rotating MP UBS solutions,
which ranges from the static marginally stable black string
to the extremal rotating marginally stable black string.

\begin{figure}[h!]
\setlength{\unitlength}{1cm}
\begin{picture}(15,8)
%\put(-1,0){\epsfig{file=Fig2l0.5new.eps,width=8cm}}
\put(-5,-4){
\includegraphics[width=160mm,angle=0,keepaspectratio]{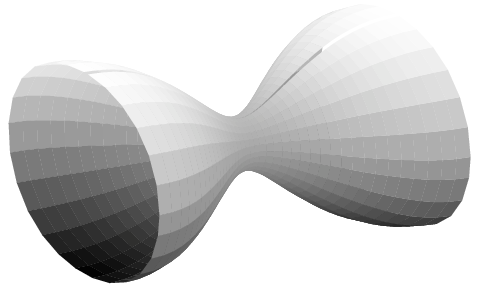}
}
\put(3,-4){
\includegraphics[width=160mm,angle=0,keepaspectratio]{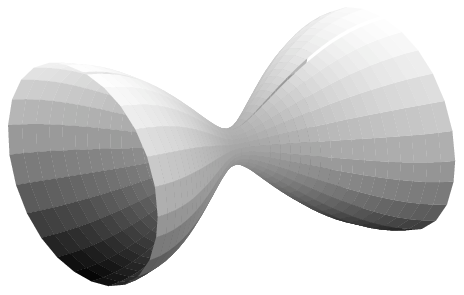}
}
\put(-5,0){
\includegraphics[width=160mm,angle=0,keepaspectratio]{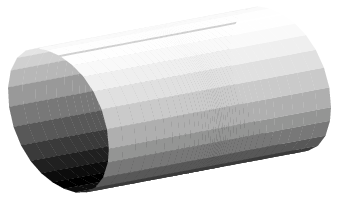}
}
\put(3,0){
\includegraphics[width=160mm,angle=0,keepaspectratio]{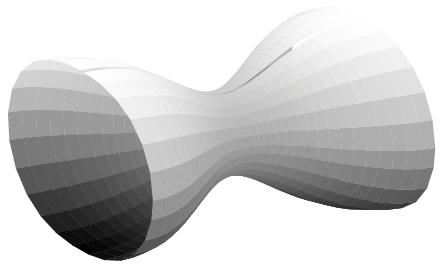}
}
\end{picture}
%{\small {\bf Figure 3.}
\caption{
The spatial embedding of the horizon
of $D=6$ rotating black string solutions
is shown for a sequence of solutions with fixed
temperature parameter $d_0=0.6$ and varying
horizon angular velocity $\Omega$:
$\Omega=0.34908$ (upper left), $\Omega=0.25$ (upper right)
$\Omega=0.212$ (lower left) and $\Omega=0.202$ (lower right),
$\Lambda$ specifies the increasing nonuniformity of the solutions.
($r_{\rm H} =1$, $L=L^{\rm crit}=4.9516$.)
}
\label{f-11}
\end{figure}

In Fig.~\ref{f-11} we exhibit
the spatial embedding of the horizon into 3-dimensional space
for a sequence of $D=6$ rotating NUBS.
In these embeddings the symmetry directions ($\varphi_1$, $\varphi_2$)
are suppressed, and the proper circumference of the horizon is plotted
against the proper length along the compact direction,
yielding a geometrical view of both the deformation of
the horizon due to rotation and the nonuniformity of the horizon
with respect to the compact coordinate.
In the rotating NUBS
the rotation leads to a deformation of the 3-sphere of the horizon, making
it oblate w.r.t.~the planes of rotation.

For the solutions of the sequence shown in Fig.~\ref{f-11}
the temperature is kept fixed with temperature parameter $d_0=0.6$.
The first solution
corresponds to the marginally stable rotating uniform black string,
which has $\Lambda=0$ and horizon angular velocity $\Omega=0.34908$.
When the horizon angular velocity is lowered,
rotating black strings with increasing nonuniformity are obtained.
Shown are solutions with nonuniformity parameter
$\Lambda=0.83$, $1.7$ and $2.9$.
(Note, that we define the radii ${\cal R}_{\rm min}$ and
${\cal R}_{\rm max}$ now via the area of the deformed horizon $3$-sphere.)
With increasing $\Lambda$ we again expect to approach
a horizon topology changing transition,
which would now lead to rotating caged black holes.
Obtaining the respective branches of rotating caged black holes
still represents a numerical challenge.

\section{Conclusions}

We have discussed certain aspects of
black holes in higher dimensions.
In particular, we have pointed out that black holes in higher dimensions
are far less restricted than black holes in four dimensions,
since higher dimensions open up a whole new range
of possibilities for the properties of black objects,
culminating in a rich phase structure.

Our first objective was the study of charged rotating black holes
with a horizon of spherical topology.
Here we saw, that the set of Einstein-Maxwell equations
simplifies strongly in odd dimensions, when all angular momenta
have equal magnitude, due to an enhanced symmetry.
Making use of this simplification, we studied
the properties of Einstein-Maxwell-Chern-Simons black holes.
We obtained black holes, which are counterrotating,
black holes, which have a negative horizon mass,
black holes, which have a static horizon but a finite
angular momentum, black holes whose horizon rotates,
but whose angular momentum vanishes,
and even black holes, which are not static,
but have a static horizon and zero angular momentum.
Moreover, we observed nonuniqueness of black holes
with spherical horizon topology.

Our next objective was to provide evidence for a
horizon topology changing transition for black strings,
i.e., for solutions, that are not asymptotically flat,
since their background manifold is ${\cal M}^{D-1} \times S^1$.
In particular, we have constructed nonuniform black string solutions,
static as well as rotating, with increasing nonuniformity.
For these NUBS we have always observed a backbending 
of the solutions close to the expected transition point
to caged black holes,
when considered versus the relative string tension.
Extrapolating the static caged black hole branches
towards the expected respective transition point has shown 
good agreement of the physical quantities of the solutions
on both types of branches.
So far, branches of rotating caged black holes have not yet been obtained,
but similar agreement is expected.

Thus, by now there is considerable evidence for the occurrence
of horizon topology changing transitions in 
nonuniform black string--caged black hole systems.
Fundamental here is of course the presence of
the Gregory-Laflamme instability,
giving rise to the branches of nonuniform black strings
in the first place.

However, a similar instability does arise also for
asymptotically flat rotating black holes in $D \ge 6$ dimensions,
possessing only a single angular momentum, 
since then this angular momentum is not bounded.
Indeed, this instability suggested, that as in the case 
of black strings, a branch
of nonuniform rotating black holes should arise at the
marginally stable solution, termed rippled or pinched black holes.

\begin{figure}[t!]
\includegraphics[width=130mm,angle=0,keepaspectratio]{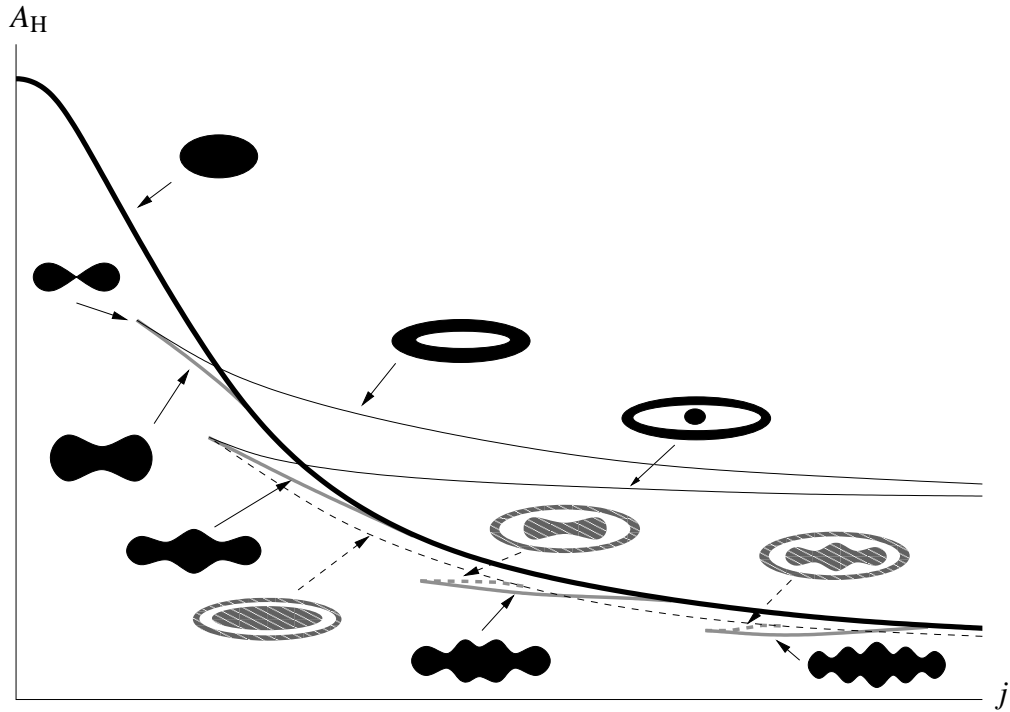}
\begin{picture}(0,0)(0,0)
\put(-1,-1){$j$}
%\put(25,250){$a_H$}
\put(-373,255){$A_{\rm H}$}
%\put(380,5){$j$}
\end{picture}
  \caption{
Proposal for the phase diagram of thermal equilibrium phases
in $D\geq 6$. The solid lines and figures have significant arguments in
their favor, while the dashed lines and figures might not exist and
admit conceivable, but more complicated, alternatives. Some
features have been drawn arbitrarily:
at any given bifurcation and in any dimension, smooth connections
are possible instead of swallowtails with cusps; also, the
bifurcation into two black Saturn phases may happen before, after, or
right at the merger with the pinched black hole. Mergers to di-rings or
multi-ring configurations that extend to asymptotically large $j$ seem
unlikely. If thermal equilibrium is not imposed, the whole semi-infinite
strip $0<A_{\rm H} <A_{\rm H}(j=0)$, $0\leq j<\infty$ is covered, and multi-rings
are possible (courtesy of Emparan et al. \cite{Emparan:2007wm}).
}
\label{f-12}
\end{figure}

Recently, this analogy has been explored further \cite{Emparan:2007wm},
culminating in the conjectured phase diagram 
for fast rotating black objects, exhibited in Fig.~\ref{f-12}.
The solid line here is the branch of MP black holes with
a single angular momentum.
As the first instability arises, a branch of pinched rotating
black holes appears. These then become increasingly deformed,
until the horizon touches itself at a horizon topology changing
solution, where it merges with the corresponding branch of black rings.
As the next instabilities arise, branches of pinched rotating black holes
with more complex nonuniform structure are conjectured to appear,
all expected to lead to horizon topology changing transitions.
Construction of these solutions and thus verification
of this phase diagram, however, represents a major challenge.

\begin{theacknowledgments}
BK gratefully acknowledges support by the German Aerospace Center,
and FNL support by the Ministerio de Educaci\'on y Ciencia.
\end{theacknowledgments}


\begin{thebibliography}{000}

%\cite{Hawking:1973uf}
\bibitem{Hawking:1973uf}
  S.~W.~Hawking and G.~F.~R.~Ellis,
  ``The Large scale structure of space-time,''
%\href{http://www.slac.stanford.edu/spires/find/hep/www?irn=6991262}{SPIRES entry}
{\it  Cambridge University Press, Cambridge, 1973}

%\cite{Friedman:1993ty}
\bibitem{Friedman:1993ty}
  J.~L.~Friedman, K.~Schleich and D.~M.~Witt,
  %``Topological censorship,''
  Phys.\ Rev.\ Lett.\  {\bf 71}, 1486 (1993)
  [Erratum-ibid.\  {\bf 75}, 1872 (1995)]
  [arXiv:gr-qc/9305017].
  %%CITATION = PRLTA,71,1486;%%

%\cite{Israel:1967za}
\bibitem{Israel:1967za}
  W.~Israel,
  %``Event horizons in static electrovac space-times,''
  Commun.\ Math.\ Phys.\  {\bf 8}, 245 (1968).
  %%CITATION = CMPHA,8,245;%%

%\cite{Robinson:1975bv}
\bibitem{Robinson:1975bv}
  D.~C.~Robinson,
  %``Uniqueness of the Kerr black hole,''
  Phys.\ Rev.\ Lett.\  {\bf 34} (1975) 905.
  %%CITATION = PRLTA,34,905;%%

%\cite{Mazur:1982db}
\bibitem{Mazur:1982db}
  P.~O.~Mazur,
  %``Proof Of Uniqueness Of The Kerr-Newman Black Hole Solution,''
  J.\ Phys.\ A  {\bf 15} (1982) 3173.
  %%CITATION = JPAGB,A15,3173;%%

%\cite{Heusler:1996}
\bibitem{Heusler:1996}
M.~Heusler,
 ``Black Hole Uniqueness Theorems,''
%Cambridge Lecture Notes in Physics
{\it Cambrigde University Press, Cambridge, 1996}

%\cite{Wald:1993ki}
\bibitem{Wald:1993ki}
  R.~M.~Wald,
  %``The First Law Of Black Hole Mechanics,''
{\it Directions in General Relativity,
%An International Symposium in Honor of the 60th Birthdays 
%of Dieter Brill and Charles Misner, College Park, Maryland, 27-29 May 1993.
%In *College Park 1993, Directions in general relativity, 
vol. 1, 358}, 
arXiv:gr-qc/9305022.
  %%CITATION = GR-QC/9305022;%%

%\cite{Smarr:1972kt}
\bibitem{Smarr:1972kt}
  L.~Smarr,
  %``Mass Formula For Kerr Black Holes,''
  Phys.\ Rev.\ Lett.\  {\bf 30}, 71 (1973).
  %%CITATION = PRLTA,30,71;%%


%\cite{Tangherlini:1963bw}
\bibitem{Tangherlini:1963bw}
  F.~R.~Tangherlini,
  %``Schwarzschild field in n dimensions and the dimensionality of space
  %problem,''
  Nuovo Cim.\  {\bf 27}, 636 (1963).
  %%CITATION = NUCIA,27,636;%%

%\cite{Myers:1986un}
\bibitem{Myers:1986un}
  R.~C.~Myers and M.~J.~Perry,
  %``Black Holes In Higher Dimensional Space-Times,''
  Annals Phys.\  {\bf 172}, 304 (1986).
  %%CITATION = APNYA,172,304;%%

%\cite{Emparan:2003sy}
\bibitem{Emparan:2003sy}
  R.~Emparan and R.~C.~Myers,
  %``Instability of ultra-spinning black holes,''
  JHEP {\bf 0309}, 025 (2003)
  [arXiv:hep-th/0308056].
  %%CITATION = JHEPA,0309,025;%%

%\cite{Emparan:2007wm}
\bibitem{Emparan:2007wm}
  R.~Emparan, T.~Harmark, V.~Niarchos, N.~A.~Obers and M.~J.~Rodriguez,
  %``The Phase Structure of Higher-Dimensional Black Rings and Black Holes,''
  arXiv:0708.2181 [hep-th].
  %%CITATION = ARXIV:0708.2181;%%

%\cite{Horowitz:1995tm}
\bibitem{Horowitz:1995tm}
  G.~T.~Horowitz and A.~Sen,
  %``Rotating Black Holes which Saturate a Bogomol'nyi Bound,''
  Phys.\ Rev.\  D {\bf 53}, 808 (1996)
  [arXiv:hep-th/9509108].
  %%CITATION = PHRVA,D53,808;%%

%\cite{Youm:1997hw}
\bibitem{Youm:1997hw}
  D.~Youm,
  %``Black holes and solitons in string theory,''
  Phys.\ Rept.\  {\bf 316}, 1 (1999)
  [arXiv:hep-th/9710046].
  %%CITATION = PRPLC,316,1;%%

%\cite{Hassan:1991mq}
\bibitem{Hassan:1991mq}
  S.~F.~Hassan and A.~Sen,
  %``Twisting classical solutions in heterotic string theory,''
  Nucl.\ Phys.\  B {\bf 375}, 103 (1992)
  [arXiv:hep-th/9109038].
  %%CITATION = NUPHA,B375,103;%%

%\cite{Sen:1994eb}
\bibitem{Sen:1994eb}
  A.~Sen,
  %``Black Hole Solutions In Heterotic String Theory On A Torus,''
  Nucl.\ Phys.\  B {\bf 440}, 421 (1995)
  [arXiv:hep-th/9411187].
  %%CITATION = NUPHA,B440,421;%%

%\cite{Llatas:1996gh}
\bibitem{Llatas:1996gh}
  P.~M.~Llatas,
  %``Electrically Charged Black-holes for the Heterotic String Compactified on a
  %$(10-D)$-torus,''
  Phys.\ Lett.\  B {\bf 397}, 63 (1997)
  [arXiv:hep-th/9605058].
  %%CITATION = PHLTA,B397,63;%%

%\cite{Kunz:2006jd}
\bibitem{Kunz:2006jd}
  J.~Kunz, D.~Maison, F.~Navarro-Lerida and J.~Viebahn,
  %``Rotating Einstein-Maxwell-dilaton black holes in D dimensions,''
  Phys.\ Lett.\  B {\bf 639}, 95 (2006)
  [arXiv:hep-th/0606005].
  %%CITATION = PHLTA,B639,95;%%

%\cite{Breckenridge:1996sn}
\bibitem{Breckenridge:1996sn}
  J.~C.~Breckenridge, D.~A.~Lowe, R.~C.~Myers, A.~W.~Peet, A.~Strominger
  and C.~Vafa,
 %``Macroscopic and Microscopic Entropy of Near-Extremal Spinning Black
 %Holes,''
  Phys.\ Lett.\ B {\bf 381}, 423 (1996)
  [arXiv:hep-th/9603078].
  %%CITATION = HEP-TH 9603078;%%

 %\cite{Breckenridge:1996is}
\bibitem{Breckenridge:1996is}
  J.~C.~Breckenridge, R.~C.~Myers, A.~W.~Peet and C.~Vafa,
 %``D-branes and spinning black holes,''
  Phys.\ Lett.\ B {\bf 391}, 93 (1997)
  [arXiv:hep-th/9602065].
  %%CITATION = HEP-TH 9602065;%%

%\cite{Cvetic:2004hs}
\bibitem{Cvetic:2004hs}
  M.~Cvetic, H.~Lu and C.~N.~Pope,
  %``Charged Kerr-de Sitter black holes in five dimensions,''
  Phys.\ Lett.\  B {\bf 598}, 273 (2004)
  [arXiv:hep-th/0406196].
  %%CITATION = PHLTA,B598,273;%%

%\cite{Chong:2005hr}
\bibitem{Chong:2005hr}
  Z.~W.~Chong, M.~Cvetic, H.~Lu and C.~N.~Pope,
  %``General non-extremal rotating black holes in minimal five-dimensional
  %gauged supergravity,''
  Phys.\ Rev.\ Lett.\  {\bf 95}, 161301 (2005)
  [arXiv:hep-th/0506029].
  %%CITATION = PRLTA,95,161301;%%

%\cite{Emparan:2001wn}
\bibitem{Emparan:2001wn}
  R.~Emparan and H.~S.~Reall,
  %``A rotating black ring in five dimensions,''
  Phys.\ Rev.\ Lett.\  {\bf 88}, 101101 (2002)
  [arXiv:hep-th/0110260].
  %%CITATION = PRLTA,88,101101;%%

%\cite{Emparan:2006mm}
\bibitem{Emparan:2006mm}
  R.~Emparan and H.~S.~Reall,
  %``Black rings,''
  Class.\ Quant.\ Grav.\  {\bf 23}, R169 (2006)
  [arXiv:hep-th/0608012].
  %%CITATION = CQGRD,23,R169;%%

%\cite{Gauntlett:2004wh}
\bibitem{Gauntlett:2004wh}
  J.~P.~Gauntlett and J.~B.~Gutowski,
  %``Concentric black rings,''
  Phys.\ Rev.\  D {\bf 71}, 025013 (2005)
  [arXiv:hep-th/0408010].
  %%CITATION = PHRVA,D71,025013;%%

%\cite{Elvang:2007rd}
\bibitem{Elvang:2007rd}
  H.~Elvang and P.~Figueras,
  %``Black Saturn,''
  JHEP {\bf 0705}, 050 (2007)
  [arXiv:hep-th/0701035].
  %%CITATION = JHEPA,0705,050;%%

%\cite{Kol:2003if}
\bibitem{Kol:2003if}
  B.~Kol, E.~Sorkin and T.~Piran,
  %``Caged black holes: Black holes in compactified spacetimes. I: Theory,''
  Phys.\ Rev.\  D {\bf 69}, 064031 (2004)
  [arXiv:hep-th/0309190].
  %%CITATION = PHRVA,D69,064031;%%

%\cite{Sorkin:2003ka}
\bibitem{Sorkin:2003ka}
  E.~Sorkin, B.~Kol and T.~Piran,
  %``Caged black holes: Black holes in compactified spacetimes. II: 5d
  %numerical implementation,''
  Phys.\ Rev.\  D {\bf 69}, 064032 (2004)
  [arXiv:hep-th/0310096].
  %%CITATION = PHRVA,D69,064032;%%

%\cite{Kudoh:2003ki}
\bibitem{Kudoh:2003ki}
  H.~Kudoh and T.~Wiseman,
  %``Properties of Kaluza-Klein black holes,''
  Prog.\ Theor.\ Phys.\  {\bf 111}, 475 (2004)
  [arXiv:hep-th/0310104].

%\cite{Harmark:2003yz}
\bibitem{Harmark:2003yz}
  T.~Harmark,
  %``Small black holes on cylinders,''
  Phys.\ Rev.\  D {\bf 69}, 104015 (2004)
  [arXiv:hep-th/0310259].
  %%CITATION = PHRVA,D69,104015;%%
  %%CITATION = PTPKA,111,475;%%

%\cite{Kol:2002xz}
\bibitem{Kol:2002xz}
  B.~Kol,
  %``Topology change in general relativity and the black-hole black-string
  %transition,''
  JHEP {\bf 0510}, 049 (2005)
  [arXiv:hep-th/0206220].
  %%CITATION = JHEPA,0510,049;%%

%\cite{Kol:2004ww}
\bibitem{Kol:2004ww}
  B.~Kol,
  %``The phase transition between caged black holes and black strings: A
  %review,''
  Phys.\ Rept.\  {\bf 422}, 119 (2006)
  [arXiv:hep-th/0411240].
  %%CITATION = PRPLC,422,119;%%

%\cite{Harmark:2005pp}
\bibitem{Harmark:2005pp}
  T.~Harmark and N.~A.~Obers,
  %``Phases of Kaluza-Klein black holes: A brief review,''
  arXiv:hep-th/0503020.
  %%CITATION = HEP-TH/0503020;%%

%\cite{Harmark:2007md}
\bibitem{Harmark:2007md}
  T.~Harmark, V.~Niarchos and N.~A.~Obers,
  %``Instabilities of black strings and branes,''
  Class.\ Quant.\ Grav.\  {\bf 24}, R1 (2007)
  [arXiv:hep-th/0701022].
  %%CITATION = CQGRD,24,R1;%%

%\cite{Gubser:2001ac}
\bibitem{Gubser:2001ac}
  S.~S.~Gubser,
  %``On non-uniform black branes,''
  Class.\ Quant.\ Grav.\  {\bf 19}, 4825 (2002)
  [arXiv:hep-th/0110193].
  %%CITATION = HEP-TH 0110193;%%

%\cite{Wiseman:2002zc}
\bibitem{Wiseman:2002zc}
  T.~Wiseman,
  %``Static axisymmetric vacuum solutions and non-uniform black strings,''
  Class.\ Quant.\ Grav.\  {\bf 20}, 1137 (2003)
  [arXiv:hep-th/0209051].
  %%CITATION = HEP-TH 0209051;%%

%\cite{Gregory:1993vy}
\bibitem{Gregory:1993vy}
  R.~Gregory and R.~Laflamme,
  %``Black strings and p-branes are unstable,''
  Phys.\ Rev.\ Lett.\  {\bf 70}, 2837 (1993)
  [arXiv:hep-th/9301052].

%\cite{Gauntlett:1998fz}
\bibitem{Gauntlett:1998fz}
  J.~P.~Gauntlett, R.~C.~Myers and P.~K.~Townsend,
 %``Black holes of D = 5 supergravity,''
  Class.\ Quant.\ Grav.\  {\bf 16}, 1 (1999)
  [arXiv:hep-th/9810204].
  %%CITATION = HEP-TH 9810204;%%

%\cite{Kunz:2005nm}
\bibitem{Kunz:2005nm}
  J.~Kunz, F.~Navarro-Lerida and A.~K.~Petersen,
  %``Five-dimensional charged rotating black holes,''
  Phys.\ Lett.\ B {\bf 614}, 104 (2005)
  [arXiv:gr-qc/0503010].
  %%CITATION = GR-QC 0503010;%%

%\cite{Kunz:2006eh}
\bibitem{Kunz:2006eh}
  J.~Kunz, F.~Navarro-Lerida and J.~Viebahn,
  %``Charged rotating black holes in odd dimensions,''
  Phys.\ Lett.\  B {\bf 639}, 362 (2006)
  [arXiv:hep-th/0605075].
  %%CITATION = PHLTA,B639,362;%%

%\cite{Kunz:2005ei}
\bibitem{Kunz:2005ei}
  J.~Kunz and F.~Navarro-Lerida,
  %``D = 5 Einstein-Maxwell-Chern-Simons black holes,''
  Phys.\ Rev.\ Lett.\  {\bf 96}, 081101 (2006)
  [arXiv:hep-th/0510250].
  %%CITATION = PRLTA,96,081101;%%

%\cite{Kunz:2006yp}
\bibitem{Kunz:2006yp}
  J.~Kunz and F.~Navarro-Lerida,
  %``Negative horizon mass for rotating black holes,''
  Phys.\ Lett.\  B {\bf 643}, 55 (2006)
  [arXiv:hep-th/0610036].
  %%CITATION = PHLTA,B643,55;%%

%\cite{Kunz:2006xk}
\bibitem{Kunz:2006xk}
  J.~Kunz and F.~Navarro-Lerida,
  %``Non-uniqueness, counterrotation, and negative horizon mass of
  %Einstein-Maxwell-Chern-Simons black holes,''
  Mod.\ Phys.\ Lett.\  A {\bf 21}, 2621 (2006)
  [arXiv:hep-th/0610075].
  %%CITATION = MPLAE,A21,2621;%%

%\cite{Kleihaus:2000kg}
\bibitem{Kleihaus:2000kg}
  B.~Kleihaus and J.~Kunz,
  %``Rotating hairy black holes,''
  Phys.\ Rev.\ Lett.\  {\bf 86}, 3704 (2001)
  [arXiv:gr-qc/0012081].
  %%CITATION = PRLTA,86,3704;%%

%\cite{Kleihaus:2002ee}
\bibitem{Kleihaus:2002ee}
  B.~Kleihaus, J.~Kunz and F.~Navarro-Lerida,
  %``Rotating Einstein-Yang-Mills black holes,''
  Phys.\ Rev.\  D {\bf 66}, 104001 (2002)
  [arXiv:gr-qc/0207042].
  %%CITATION = PHRVA,D66,104001;%%

%\cite{Kleihaus:2002tc}
\bibitem{Kleihaus:2002tc}
  B.~Kleihaus, J.~Kunz and F.~Navarro-Lerida,
  %``Global charges of stationary non-Abelian black holes,''
  Phys.\ Rev.\ Lett.\  {\bf 90}, 171101 (2003)
  [arXiv:hep-th/0210197].
  %%CITATION = PRLTA,90,171101;%%

%\cite{Kunz:2007jq}
\bibitem{Kunz:2007jq}
  J.~Kunz, F.~Navarro-Lerida and E.~Radu,
  %``Higher dimensional rotating black holes in Einstein-Maxwell theory with
  %negative cosmological constant,''
  Phys.\ Lett.\  B {\bf 649}, 463 (2007)
  [arXiv:gr-qc/0702086].
  %%CITATION = PHLTA,B649,463;%%

%\cite{Brihaye:2007bi}
\bibitem{Brihaye:2007bi}
  Y.~Brihaye and T.~Delsate,
  %``Charged-rotating black holes and black strings in higher dimensional
  %Einstein-Maxwell theory with a positive cosmological constant,''
  Class.\ Quant.\ Grav.\  {\bf 24}, 4691 (2007)
  [arXiv:gr-qc/0703146].
  %%CITATION = CQGRD,24,4691;%%

%\cite{Aliev:2004ec}
\bibitem{Aliev:2004ec}
  A.~N.~Aliev and V.~P.~Frolov,
  %``Five dimensional rotating black hole in a uniform magnetic field: The
  %gyromagnetic ratio,''
  Phys.\ Rev.\  D {\bf 69}, 084022 (2004)
  [arXiv:hep-th/0401095].
  %%CITATION = PHRVA,D69,084022;%%

%\cite{Aliev:2006yk}
\bibitem{Aliev:2006yk}
  A.~N.~Aliev,
  %``Rotating black holes in higher dimensional Einstein-Maxwell gravity,''
  Phys.\ Rev.\  D {\bf 74}, 024011 (2006)
  [arXiv:hep-th/0604207].
  %%CITATION = PHRVA,D74,024011;%%

%\cite{NavarroLerida:2007ez}
\bibitem{NavarroLerida:2007ez}
  F.~Navarro-Lerida,
  %``Perturbative Charged Rotating 5D Einstein-Maxwell Black Holes,''
  arXiv:0706.0591 [hep-th].
  %%CITATION = ARXIV:0706.0591;%%

%\cite{Gibbons:1993xt}
\bibitem{Gibbons:1993xt}
  G.~W.~Gibbons, D.~Kastor, L.~A.~J.~London, P.~K.~Townsend and J.~H.~Traschen,
  %``Supersymmetric selfgravitating solitons,''
  Nucl.\ Phys.\  B {\bf 416}, 850 (1994)
  [arXiv:hep-th/9310118].
  %%CITATION = NUPHA,B416,850;%%

%\cite{Herdeiro:2000ap}
\bibitem{Herdeiro:2000ap}
  C.~A.~R.~Herdeiro,
  %``Special properties of five dimensional BPS rotating black holes,''
  Nucl.\ Phys.\  B {\bf 582}, 363 (2000)
  [arXiv:hep-th/0003063].
  %%CITATION = NUPHA,B582,363;%%

%\cite{Townsend:2002yf}
\bibitem{Townsend:2002yf}
  P.~K.~Townsend,
  %``Surprises with angular momentum,''
  Annales Henri Poincare {\bf 4}, S183 (2003)
  [arXiv:hep-th/0211008].
  %%CITATION = AHPJF,4,S183;%%

%\cite{Kleihaus:2003df}
\bibitem{Kleihaus:2003df}
  B.~Kleihaus, J.~Kunz and F.~Navarro-Lerida,
  %``Stationary black holes with static and counterrotating horizons,''
  Phys.\ Rev.\  D {\bf 69}, 081501 (2004)
  [arXiv:gr-qc/0309082].
  %%CITATION = PHRVA,D69,081501;%%

%\cite{Brodbeck:1997ek}
\bibitem{Brodbeck:1997ek}
  O.~Brodbeck, M.~Heusler, N.~Straumann and M.~S.~Volkov,
  %``Rotating solitons and nonrotating, nonstatic black holes,''
  Phys.\ Rev.\ Lett.\  {\bf 79}, 4310 (1997)
  [arXiv:gr-qc/9707057].
  %%CITATION = PRLTA,79,4310;%%

%\cite{Kleihaus:2006ee}
\bibitem{Kleihaus:2006ee}
  B.~Kleihaus, J.~Kunz and E.~Radu,
  %``New nonuniform black string solutions,''
  JHEP {\bf 0606}, 016 (2006)
  [arXiv:hep-th/0603119].
  %%CITATION = JHEPA,0606,016;%%

%\cite{Traschen:2001pb}
\bibitem{Traschen:2001pb}
  J.~H.~Traschen and D.~Fox,
  %``Tension perturbations of black brane spacetimes,''
  Class.\ Quant.\ Grav.\  {\bf 21}, 289 (2004)
  [arXiv:gr-qc/0103106].
  %%CITATION = CQGRD,21,289;%%

%\cite{Harmark:2003dg}
\bibitem{Harmark:2003dg}
  T.~Harmark and N.~A.~Obers,
  %``New phase diagram for black holes and strings on cylinders,''
  Class.\ Quant.\ Grav.\  {\bf 21}, 1709 (2004)
  [arXiv:hep-th/0309116].
  %%CITATION = CQGRD,21,1709;%%

%\cite{Wiseman:2002ti}
\bibitem{Wiseman:2002ti}
  T.~Wiseman,
  %``From black strings to black holes,''
  Class.\ Quant.\ Grav.\  {\bf 20}, 1177 (2003)
  [arXiv:hep-th/0211028].
  %%CITATION = HEP-TH 0211028;%%

%\cite{Kol:2003ja}
\bibitem{Kol:2003ja}
  B.~Kol and T.~Wiseman,
  %``Evidence that highly non-uniform black strings have a conical waist,''
  Class.\ Quant.\ Grav.\  {\bf 20}, 3493 (2003)
  [arXiv:hep-th/0304070].
  %%CITATION = HEP-TH 0304070;%%

%\cite{Kudoh:2004hs}
\bibitem{Kudoh:2004hs}
  H.~Kudoh and T.~Wiseman,
  %``Connecting black holes and black strings,''
  Phys.\ Rev.\ Lett.\  {\bf 94}, 161102 (2005)
  [arXiv:hep-th/0409111].
  %%CITATION = PRLTA,94,161102;%%

%\cite{Kleihaus:2007dg}
\bibitem{Kleihaus:2007dg}
  B.~Kleihaus, J.~Kunz and E.~Radu,
  %``Rotating nonuniform black string solutions,''
  JHEP {\bf 0705}, 058 (2007)
  [arXiv:hep-th/0702053].
  %%CITATION = JHEPA,0705,058;%%

%\cite{Kleihaus:2007cf}
\bibitem{Kleihaus:2007cf}
  B.~Kleihaus and J.~Kunz,
  %``Interior of Nonuniform Black Strings,''
  arXiv:0710.1726 [hep-th].
  %%CITATION = ARXIV:0710.1726;%%

\end{thebibliography}
\end{document}